\documentclass{aa}  
\usepackage{graphicx}
\usepackage{txfonts}
\begin{document}

   \title{A deep X-ray view of the bare AGN 
     Ark120.}

   \subtitle{V. Spin determination from disc-Comptonisation efficiency method}

   \author{D.\ Porquet\inst{1}   
           \and C.\ Done\inst{2} 
            \and J.\ N.\ Reeves\inst{3}
             \and N.\ Grosso\inst{1} 
           \and A.\ Marinucci\inst{4}  
          \and G.\ Matt\inst{4}
          \and A.\ Lobban\inst{5} 
          \and E.\ Nardini\inst{6}
          \and V.\ Braito\inst{7}
          \and F.\ Marin\inst{8} 
        \and A.\ Kubota\inst{9}
        \and C.\ Ricci\inst{10,11}
         \and M.\ Koss\inst{12,13}
        \and D.\ Stern\inst{14}
         \and D.\ Ballantyne\inst{15}
         \and D.\ Farrah\inst{16}
 }
   \institute{Aix-Marseille Univ, CNRS, CNES, LAM, Marseille, France   \email{delphine.porquet@lam.fr}
              \and Department of Physics, University of Durham, South Road, Durham DH1 3LE, UK
  \and Center for Space Science and Technology, University of Maryland
Baltimore County, 1000 Hilltop Circle, Baltimore, MD 21250, USA
             \and Dipartimento di Matematica e Fisica, Università degli Studi Roma Tre, via della Vasca Navale 84, I-00146 Roma, Italy
             \and Astrophysics Group, School of Physical and Geographical Sciences, Keele University, Keele, Staffordshire, ST5 5BG, UK
           \and INAF - Osservatorio Astrofisico di Arcetri, Largo Enrico Fermi 5, I-50125 Firenze, Italy 
           \and INAF - Osservatorio Astronomico di Brera, Via Bianchi 46, I-23807 Merate (LC), Italy 
  \and  Universit{\'e} de Strasbourg, CNRS, Observatoire
     astronomique de Strasbourg, UMR 7550, F-67000 Strasbourg, France
     \and Department of Electronic Information Systems, Shibaura Institute of Technology, 307 Fukasaku, Minuma-ku, Saitama-shi, Saitam 337-8570, Japan
     \and N\'ucleo de Astronom\'ia de la Facultad de Ingenier\'ia, Universidad Diego Portales, Av. Ej\'ercito Libertador 441, Santiago, Chile
     \and Kavli Institute for Astronomy and Astrophysics, Peking University, Beijing 100871, China
     \and Institute for Particle Physics and Astrophysics, ETH Zurich, Wolfgang-Pauli-Strasse 27, CH-8093 Zürich, Switzerland     
     \and Eureka Scientific Inc., 2452 Delmer Street, Suite 100, Oakland, CA 94602, USA
     \and Jet Propulsion Laboratory, California Institute of Technology, 4800 Oak Grove Drive, Mail Stop 169-221, Pasadena, CA 91109, USA
     \and Center for Relativistic Astrophysics, School of Physics, Georgia Institute of Technology, 837 State Street, Atlanta, GA 30332-0430, USA
     \and Department of Physics, Virginia Tech, Blacksburg, VA 24061, USA
     }

   \date{Received , 2018; accepted , 2018}

 
  \abstract
   {
   The spin of supermassive black holes (SMBH) in active galactic nuclei (AGN) can be determined from spectral signature(s) of relativistic reflection such as the X-ray iron K$\alpha$ line profile, but 
this can be rather uncertain when the line of sight intersects the so-called warm absorber and/or other
wind components as these distort the continuum shape. 
    Therefore, AGN showing no (or very weak) 
     intrinsic absorption along the line-of-sight such as Ark\,120, a so-called bare AGN, are the
     ideal targets for SMBH spin measurements. 
    However, in our previous work on Ark 120, we found that its 2014 X-ray spectrum is dominated by Comptonisation, while the relativistic reflection emission only originates at tens of gravitational radii from the SMBH. As a result, we could not constrain the SMBH spin from disc reflection alone.
    }
   {
   Our aim is to determine the SMBH spin in Ark\,120 from an alternative technique based on the global energetics of the disc-corona system. 
   Indeed, the mass accretion rate ($\dot{M}$) through the outer disc can be measured from the optical-UV emission, 
   while the bolometric luminosity ($L_{\rm bol}$) can be fairly well constrained from the optical to hard X-rays spectral energy distribution, giving access to the accretion efficiency
   $\eta=L_{\rm bol}/(\dot{M}c^2)$ which depends on the SMBH spin. 
   }
   {
   The spectral analysis uses simultaneous {\sl XMM-Newton} (OM and pn) and {\sl
       NuSTAR} observations on 2014 March 22 and
      2013 February 18. We applied the {\sc optxconv} model (based on 
     {\sc optxagnf}) to self consistently reproduce the emission from the inner corona (warm and hot thermal Comptonisation) and the outer disc (colour temperature corrected black body),  
     taking into account both the disc inclination angle and relativistic effects. 
    For self-consistency, we modelled the mild relativistic reflection of the incident Comptonisation components using the {\sc xilconv} convolution model.
}
   {
   We infer a SMBH spin of 0.83$^{+0.05}_{-0.03}$, adopting the SMBH  reverberation mass of 1.50$\times$10$^{8}$\,M$_{\odot}$.  
   In addition, we find that the coronal radius decreases with increasing flux (by about a factor of two), from 
   85$^{+13}_{-10}$\,$R_{\rm g}$ in 2013 to 14$\pm$3\,$R_{\rm g}$ in 2014. 
     } 
   { This is the first time that such a constraint is obtained for a
   SMBH spin from this technique, thanks to
   the bare properties of Ark\,120, its well determined SMBH reverberation mass, 
   and the presence of a mild relativistic reflection component in 2014 which allows us to constrain the disc inclination angle.  
   We caution that these results
   depend on the detailed disc-corona structure, 
   which is not yet fully established.
   However, the realistic parameter values (e.g. $L_{\rm bol}/L_{\rm Edd}$, disc inclination angle) 
   found suggest that this is a promising method  
   to determine spin in moderate-$\dot{M}$ AGN.
}
   \maketitle
%

\section{Introduction}

In the standard paradigm (the so-called no hair theorem), astrophysical black holes (BH) are
described by their mass and their angular momentum commonly called
spin. The spin is usually expressed in terms of the dimensionless parameter
$a\equiv cJ/(GM_{\rm BH}^{2})$, where $c$, $J$, $G$ and $M_{\rm BH}$
are the speed of light, the angular momentum, the  Gravitational constant and the black hole mass, respectively. In stellar-mass BHs (black hole X-ray binaries, BHXBs), the spin is expected
to be native \citep[][but see \citealt{Fragos15}]{King99}; while, in
SMBHs (with masses spanning from a few millions to several 
billions solar masses), the spin is related to the accretion-ejection history of SMBHs, for example chaotic versus coherent accretion, relativistic jets, and to the galaxy merger history \citep[e.g.][]{Blandford77,Berti08,King08}.

For BHXBs, there are up-to-now four main
 methods that can be applied to X-ray data to
determine their spin \citep[e.g.][and references therein]{Remillard06,MillerJ09,McClintock11,Reynolds14}: Spectral fitting of the relativistic reflection iron K$\alpha$ line profile, spectral fitting of the thermal continuum emission (also called
'disc continuum fitting' method), quasi periodic oscillations (QPO), 
and polarimetry. 
The first two are the most used and depend on the accretion 
disc extending down to the innermost stable circular orbit (ISCO) radius, $R_{\rm ISCO}$, which is spin dependent. 
Determining the spin from the QPO depends on the assumed model -- 
even with current data favouring Lense-Thirring precession for low frequency QPOs \citep{Ingram09,Ingram16}.  
Indeed, the constraints on the spin are not tight unless combined with models for the high frequency QPOs \citep{Motta16}
whose origin is more uncertain. 
The last technique, X-ray polarimetry \cite[e.g.][]{Dovciak04b,Schnittman09} is waiting for the launch of the next  
generation of X-ray polarimeters, such as the {\sl Imaging X-ray Polarimetry Explorer} (IXPE, a NASA Small Explorer planned for launch in 2021; \citealt{Weisskopf16}).

In the case of SMBHs in AGN, the method based on X-ray spectral analysis of the relativistic reflection signature(s) 
was the only one used until recently  \citep[e.g.][and references
therein]{Reynolds14}. Indeed, disc continuum fitting is more difficult in local AGN for two reasons. 
Firstly, whereas the disc radiates in the X-rays for BHXBs, 
the disc models predict a peak temperature for typical broad-line Seyfert 1s (BLS1s) AGN
(with a BH mass of 10$^{8}$\,M$_{\odot}$ accreting at $L_{\rm bol}$/$L_{\rm Edd}$=0.2) 
in the extreme UV ($\sim$20\,eV) that is unobservable due to Galactic absorption. 
Secondly, the observed emission in typical AGN is not generally as disc-dominated as observed in BHXBs at similar $L_{\rm bol}$/L$_{\rm Edd}$=0.2. 
Instead, the optically-thick, geometrically-thin disc
emission appears to turn over in the far UV, connecting to an upturn in the observed soft X-ray flux. This
can be fit by an additional warm Comptonisation component with $kT_e\sim 0.1-0.5$~keV, which is optically-thick with $\tau\sim 10-20$ \citep[e.g.][]{P04a,Piconcelli05,Bianchi09,Scott12,Petrucci18}, 
 very different from the standard hot X-ray corona which has $\tau\sim 1$, and $kT_{\rm e}$ $\sim$ 30--150~keV \citep[e.g.][but see for some exceptions,  
 \citealt{Matt15,Tortosa17,Kara17,TurnerJ18}]{Brenneman14,Balokovic15,Fabian15,Fabian17,Marinucci14b,Marinucci16,Tortosa18}. 
However, as pointed out by \cite{Done12}, one exception can be 
AGN with much lower BH masses and higher  $L_{\rm bol}$/$L_{\rm Edd}$ such
as the narrow-line Seyfert 1s (NLS1s) 
for which the disc emission is predicted to extend to the soft X-rays
and peak near 0.1\,keV, and where such disc-dominated spectra are often seen. 
However, even here the disc models drop much more sharply than the observed soft
X-ray shape, these also require a small additional 
warm Compton component as well as the dominant 
disc emission and weak (and steep) X-ray coronal emission typically seen in BHXBs at high Eddington ratio \citep{Done12,Jin12}. 

\cite{Done12} developed a radially stratified two-zone Comptonisation disc model (see their Figure~5), called {\sc optxagnf}. This model conserves energy, assuming that the emissivity is set by the standard geometrically thin disc \cite{Novikov73}'s relation, but the energy generated by mass accretion between $R_{\rm corona}$ and $R_{\rm ISCO}$  is dissipated as both warm and hot coronal Comptonisation emission, while the outer disc ($R$\,$>$\,$R_{\rm corona}$) emits in the optical-UV as expected for a multi-colour black-body disc. 
This model has been applied to
some disc-dominated NLS1s (that is, those where $R_{\rm corona}$ is close to $R_{\rm ISCO}$) to constrain spin and mass \citep{Done12,Jin12,Done13,Done16}; 
but typically the spin value was either fixed or almost unconstrained due to large uncertainties in the BH mass. 
For disc-dominated objects, pure disc models can also be used for spin measurements but, until now, have only provided weak constraints due to large uncertainties on the BH mass of the considered AGN, for example \object{SDSS J094533.99+100950.1}, \object{NGC 3783}, and \object{H1821+643} \citep{Czerny11b,Capellupo17}. This latter technique is similar to the disc-continuum fitting method (which also depends on BH mass, distance and disc inclination) used for BHXBs.  

The {\sc optxagnf} model can also fit the BLS1s for which the soft X-ray excess is found to produce mainly by warm Comptonisation  \citep{Done12,Jin12,Mehdipour11,Mehdipour15,Porquet18} rather than by relativistic reflection \citep{Crummy06}. 
Indeed, in case the X-ray spectrum is mainly due to warm and hot Comptonisation the spin can still be constrained from the global energetics of the flow \citep{Done12}. This was first explored by \cite{Davis11}, who  assumed that the optical-UV emission was produced in the outer disc, so that the mass accretion rate for a standard disc could be constrained simply from the optical-UV luminosity as: \\
\begin{equation}\label{eq:Lband}
L_{\rm opt-UV}\propto (M_{\rm BH} \dot{M})^{2/3} \cos \theta, 
\end{equation}
where $M_{\rm BH}$ is the SMBH mass, $\dot{M}$ the absolute accretion rate and $\theta$ the accretion disc inclination angle. 
If $M_{\rm BH}$ is known, for example from reverberation mapping, and $\theta$ is constrained or already known, then this relation determines the mass accretion rate through the outer disc. 
The bolometric luminosity is:\\
\begin{equation}\label{eq:Lbol}
L_{\rm bol}=\eta \dot{M} c^2,
\end{equation}
where $\eta$ is the accretion radiative efficiency that indicates how much binding energy at the ISCO is radiated away. Since the ISCO radius varies monotonically with the black hole spin value \citep{Bardeen72}, $\eta$ is directly related to the black hole spin, assuming that the inner radius of the accretion disc corresponds to the ISCO.  
\cite{Davis11} applied this disc-Comptonisation efficiency method to a sample of bright QSOs. They find that $\eta$ increased with $M_{\rm BH}$, being consistent with low spin for lower mass SMBH but requiring higher spin for the most massive objects in their sample. This is consistent with the results of \cite{Jin12}, where  their sample of nearby, fairly low mass SMBH could all be fit with zero spin (that is, $\eta=0.057$) models. 

\object{Ark 120} (z=0.03271; \citealt{Theureau05}) is the brightest and cleanest bare AGN known. 
Indeed, its UV and X-ray spectra are `contaminated' neither by line of sight warm absorption signatures \citep{Crenshaw99,Vaughan04,
   Reeves16} nor by a neutral intrinsic absorber \citep{Reeves16}.  
Ark\,120 is also free from  intrinsic reddening
 in its infrared-optical-UV continuum \citep{Ward87,Vasudevan09}, 
 though there are non-negligible 
UV reddening and X-ray absorption from our own Galaxy in its direction. 
Moreover, the SMBH mass of Ark\,120 is well constrained thanks to reverberation mapping measurement performed by \cite{Peterson04} who assumed the virial factor\footnote{A definition of the virial factor is given in section~\ref{sec:discussion}.} from \cite{Onken04}, and obtained 1.50$\pm$0.19$\times$10$^{8}$\,M$_{\odot}$.  
In \cite{Porquet18}, using a deep X-ray observation performed in March 2014 with {\sl XMM-Newton} and {\sl NuSTAR}, 
we show that the X-ray spectra of Ark\,120 is dominated by warm and hot Comptonisation components. 
A mild reflection component is still required above about 10\,keV, but with a low degree of relativistic smearing indicating that the relativistic reflection only occurs beyond several 10s of $R_{\rm g}$ (see also \citealt{Nardini16}), and therefore does not enable us to infer any constraint on the SMBH spin from disc reflection alone.  

In this work, we report on the Ark\,120 spectral energy distribution
(SED) fitting using the disc-Comptonisation efficiency method 
combining simultaneous optical--UV ({\sl XMM-Newton}/OM), and X-rays ({\sl XMM-Newton}/pn and 
{\sl NuSTAR}) observations performed on 2014 March 22 and 2013 February 18. 
In section,~\ref{sec:obs}, we describe the
 data reduction procedure, while the spectral modelling is described in section~\ref{sec:models}. 
 The simultaneous optical to hard X-ray data analysis for the 2014 March 22 and the 2013 February
18 observations is reported in section~\ref{sec:analysis}. 
In section~\ref{sec:discussion} our main results are discussed, followed by our conclusions in section~\ref{sec:conclusion}.

\section{Observation, data reduction}\label{sec:obs}

\begin{figure}[t!]
\includegraphics[width=1.0\columnwidth,angle=0]{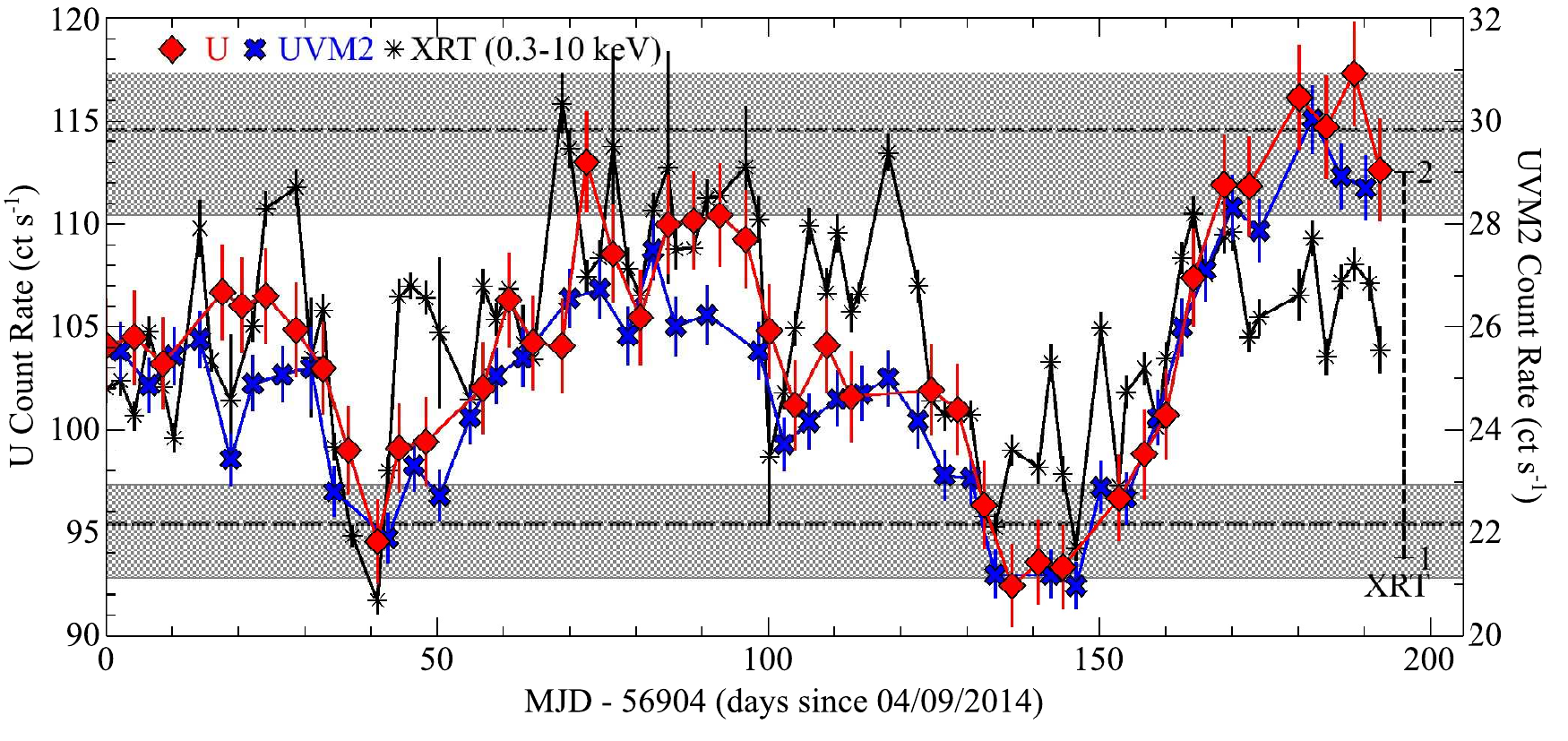} \\
\caption{The {\sl Swift} UVOT light curve of Ark 120 showing the corrected count rates
in the U (red) and UVM2 (blue) bands (adapted from \citealt{Lobban18}). 
Each point corresponds to a single image observation. 
The 0.3–10 keV XRT light curve (black) is overlaid with an additional y-axis scale. 
The lower and upper grey shaded areas correspond to the XRT count rates at the 2013 and 2014 {\sl XMM-Newton} 0.3-10\,keV fluxes, respectively. The horizontal dashed lines correspond to the mean flux for each dataset.}
\label{fig:swift}
\end{figure}

Ark 120 was observed by {\sl XMM-Newton} \citep{Jansen01} over four consecutive orbits
between 2014 March 18 and March 24 (PI: D.\ Porquet).
Here, we used the 2014 March 22 observation, which was the only one that was simultaneous with a {\sl NuSTAR} observation (PI: NuSTAR AGN team). 
We also analysed an earlier joint {\sl XMM-Newton} and {\sl NuSTAR} observation, which was performed
  in a single {\sl XMM-Newton} orbit on 2013 February 18 (PI: G.\ Matt). 
  During this 2013 observation, the X-ray flux of Ark 120 was about a factor of two lower than in 2014 
  \citep{Matt14,Marinucci19}, while the optical-UV flux was also lower \citep{Lobban18}. 
  Figure~\ref{fig:swift} shows the six-month {\sl Swift} UVOT and XRT light curves \citep{Lobban18}. We note that the shaded areas show the equivalent XRT rates,  
  corresponding to the flux levels measured from the 2013 and 2014 XMM-Newton observations described above. This illustrates that during the 2013 and 2014 {\sl XMM-Newton} observations,
   the source was observed very close to its lowest and highest flux state levels, respectively.  
   The log of the observations of Ark\,120 used in this work
is reported in Table~\ref{tab:log}. 

\begin{table*}[ht!]
\small
\caption{Observation log of the data analysed in this work for Ark\,120.}
\label{to}
\begin{tabular}{c@{\hspace{15pt}}r@{\hspace{15pt}}c@{\hspace{15pt}}c@{\hspace{15pt}}c@{\hspace{15pt}}c}
\hline \hline
Mission & Obs.\,ID & Obs.\,Start (UTC) & Exp.$^a$ & $\mathcal{C}^b$\\
        &          &                   & (ks) & (s$^{-1}$) \\
\hline
{\sl XMM-Newton} & 0721600401 & 2014 March 22 -- 08:25:17 & 82.4 & 25.23$\pm$0.02 (pn) \\
{\sl NuSTAR}  & 60001044004 & 2014 March 22 -- 09:31:07 &65.5 &1.089$\pm$0.004 (FPMA) \\
           &             &                            &65.3 &1.072$\pm$0.004 (FPMB) \\
{\sl XMM-Newton} & 0693781501  & 2013 February 18 -- 11:45:48 & 87.7 
& 10.30$\pm$0.01 (pn)  \\
{\sl NUSTAR}     &60001044002    & 2013 February 18 -- 10:46:07 & 79.5 & 0.626$\pm$0.003
(FPMA)\\
                  &    &                  & 79.4 & 0.598$\pm$0.003 FPMB)\\
\hline
\hline
\end{tabular}
\label{tab:log}
\flushleft
\small{\textit{Notes.} $^a$Net exposure in ks. $^b$Source count rate
  over the 0.3--10 keV for {\sl XMM-Newton}/pn and over 3-79\,keV for {\sl NuSTAR}.} 
\end{table*}

\subsection{{\sl XMM-Newton} data reduction}

For the data reduction we used the Science Analysis System (SAS) v16.1.0, applying the
latest calibrations available on 2018 February 2. 
This updated calibration, compared to \cite{Porquet18}, results in a better 7--10\,keV cross-calibration 
between {\sl pn} and {\sl NuSTAR} data with a slight steepening of the 2-10\,keV photon indices by about +0.03,
in other words, by about 1.6\%.

\subsubsection{pn data}

Due to the high source
brightness, the EPIC-pn camera \citep{Struder01} was operated in Small Window mode 
 to prevent any pile-up. The 2013 and 2014 pn spectra were extracted 
from circular regions centred on Ark\,120, with radii of
30${\arcsec}$. We selected the event patterns 0-4, that is,
single and double pixels, while we also applied {\sc FLAG==0} 
in order that all events at the edge of a CCD and at the edge of a bad pixel were excluded.   
The background spectra were extracted from a rectangular region in the
lower part of the small window that contains no (or negligible) source
photons. After the
correction for dead time and background flaring, the total net
pn exposures were about 82\,ks for the 2014 observation and about 88\,ks for the
2013 observation. Redistribution matrices and ancillary
response files for the two pn spectra were generated with the SAS tasks
{\sc rmfgen} and {\sc arfgen}. 
We used the time-averaged pn spectra since, as shown in \cite{Lobban18},  
the spectral variability within a single orbit is slow and
 moderate. The 0.3--10\,keV pn spectra were binned to give 100 counts per bin. 

A gain shift was applied to take into account the known inaccuracy of the EPIC-pn
energy scale likely due to inaccuracies in the long-term charge
transfer inefficiency (CTI) calibration.\footnote{http://xmm2.esac.esa.int/docs/documents/CAL-SRN-0300-1-0.pdf.} 
The {\sc gain} xspec command allows us to modify accordingly the response file gain and is
characterised by two parameters: {\sc slope} and {\sc intercept} (in units of keV). 
The new energy is calculated by $E^{\prime}$ = $E/ \langle$ {\sc slope} $\rangle$ $-$
$\langle$ {\sc intercept} $\rangle$.  
For the 2014 observations, we fit simultaneously the four available pn spectra 
tying the {\sc gain} parameters values using the following baseline
model: {\sc tbnew}$\times$({\sc comptt}+{\sc zpo}+{\sc
  zga(broad)}+{\sc 3$\times$zga(BLR)}). 
Indeed, such modelling is
adequate for the 2014 observation as shown in \cite{Porquet18}. 
The parameters of the broad Gaussian and of the three BLR Gaussian
lines were tied between the four observations. 
We allowed to vary between each
observation: kT$_{\rm e}$, $\tau$, normalisation(comptt), $\Gamma$ and
normalisation(zpo). 
We infered {\sc slope}=1.0083$^{+0.0001}_{-0.0004}$ and {\sc
  intercept}=4.77$^{+0.11}_{-0.01}\times$10$^{-3}$\,keV. 
We performed the same modelling for the 2013 pn spectrum but without the need
of a broad Gaussian line, and we infered {\sc slope}=1.0073$^{+0.0005}_{-0.0007}$  and {\sc
  intercept}=$-$1.72$^{+0.14}_{-0.46}\times$10$^{-3}$\,keV.

\subsubsection{OM data}\label{sec:OM}

We used the {\sl XMM-Newton} optical-UV Monitor telescope (hereafter OM; \citealt{Mason01}). 
For the March 2014 observation, we acquired about about five $\sim$ 1.2 ks exposures in default imaging
and fast mode consecutively through the V (effective wavelength= 5\,430\,\AA), B (4\,500\,\AA), U (3\,440\,\AA), UVW1 (2\,910\,\AA) and UVM2 (2\,310\,\AA) 
filters before spending the rest of the observation acquiring exposures
with the UVW2 (2\,120\,\AA) filter.
We did not use the (redundant) fast mode data reported in \cite{Lobban18}.
For the 2013 observation, a series of snapshots were consecutively acquired with the UVW1 (ten $\sim$3.4\,ks exposures), UVM2 (ten $\sim$4.4\,ks exposures) and UVW2 (ten $\sim$4.4\,ks exposures) filters. \\

The imaging mode data were processed using the {\tt SAS} script {\tt omichain} 
which takes into account all necessary
 calibration processes (e.g. flat-fielding), and runs a source detection
algorithm before performing aperture photometry on each detected source, 
and combines the source lists from separate exposures into a single master list  
to compute mean corrected count rates. 
The optical and UV counterparts of Ark~120 detected with the OM is point-like.
The FWHM is $1\farcs5$ and 3\arcsec\ with the V and UV filters, respectively \citep{Mason01}.
The aperture radius is 12 unbinned pixels (corresponding to $5\farcs7$ for $0\farcs$4765 square pixels) 
and the background is estimated within an annulus region: 
for the optical filters the inner and outer radii of the annulus are 14.0 and 25.1 unbinned pixels, respectively
(corresponding to $6\farcs7$ and $11\farcs9$, respectively); 
for the UV filters the inner and outer radius of the annulus are 37.0 and 42.4 unbinned pixels, respectively
(corresponding to $17\farcs6$ and $20\farcs2$, respectively). 

\begin{figure}[t]
\centering
\includegraphics[angle=0,width=\columnwidth]{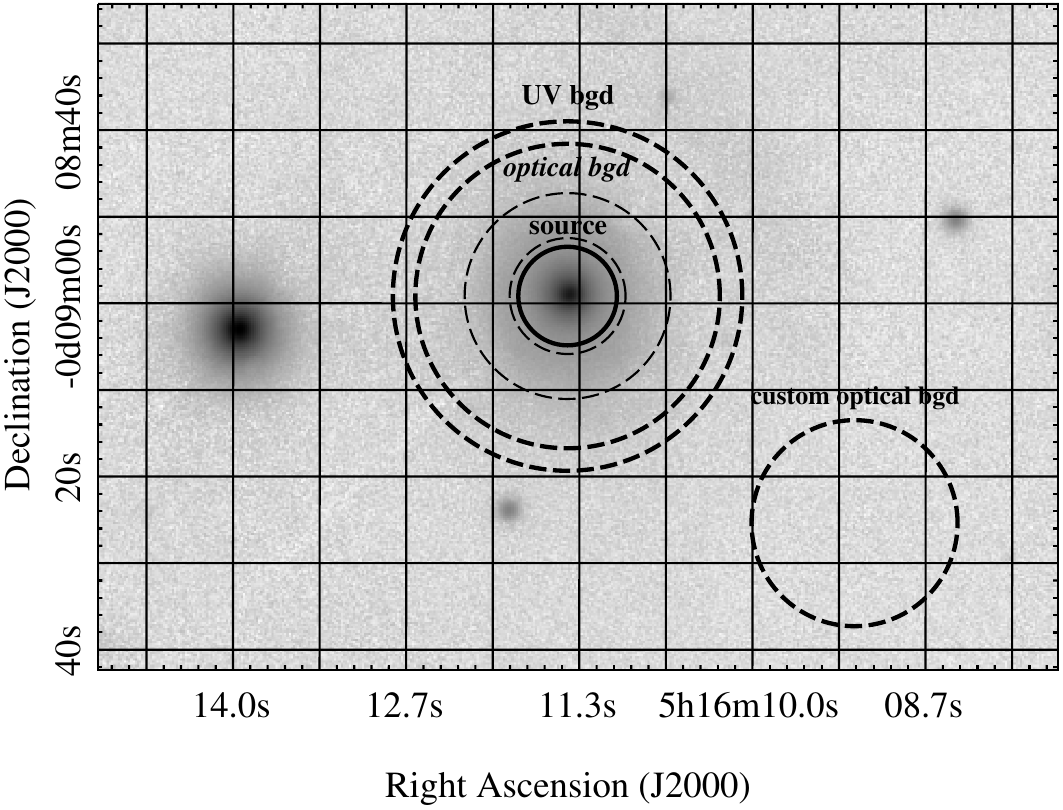}
\caption{Pan-STARRS-1 $g$-filter image of Ark~120. 
North is up and east is left.
The intensity is non-linearly scaled using an asinh transformation.
The circle and the dashed annulus are the OM source and background apertures, 
respectively. The dashed circle is the custom background aperture that we used for the optical photometry.}
\label{fig:panstarrs}
\end{figure}

For comparison we show in Fig.~\ref{fig:panstarrs} the {\tt omichain} apertures for the V-filter overlaid 
on the Pan-STARRS-1 g-filter image\footnote{The Pan-STARRS-1 image cutout server is available at 
http://ps1images.stsci.edu\,.} \citep{Chambers16}.
As the {\tt omichain} background apertures for the optical filter are located on the galaxy disc, we recomputed  
the OM optical photometry with the {\tt SAS} task {\tt omsource} 
from the unbinned central images in the detector plane, corrected for the modulo-8 pattern (*OM*IMAGE$\_$0000.FIT) 
using a background aperture of $\sim12\arcsec$-radius located $\sim43\arcsec$ SW of Ark\,120.

\begin{figure}[t]
\centering
\includegraphics[angle=0,width=\columnwidth]{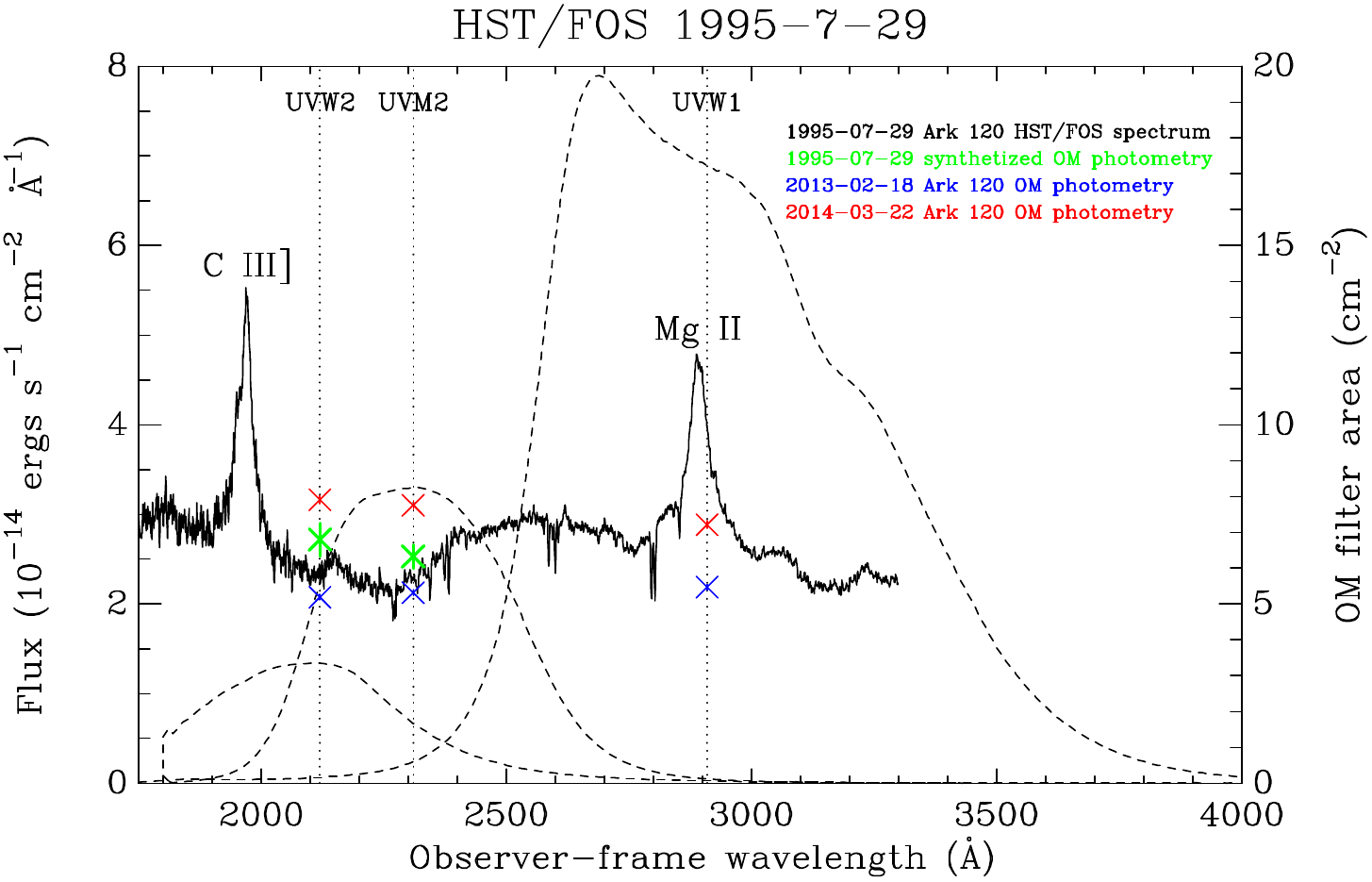}
\caption{Wavelength coverage of the UV emission of Ark\,120 with the OM broad-band UV-filters. The continuous line is the observed UV spectrum of Ark~120 obtained with the Hubble Space Telescope Faint Object Spectrograph (HST/FOS) on 1995 July 29 \citep{kuraszkiewicz04}, where the main broad emission lines are labelled. The faint absorption lines are galactic in origin \citep{Crenshaw99}. The dashed lines are the OM filter areas. The green data are the synthetised photometry that we computed from this HST/FOS spectrum using the OM broad-band UV-filter profiles. The red and blue  data are the observed mean UVW2, UVM2, and UVW1 fluxes from the corrected observed mean count rates on 2014 March 22 and on 2013 February 18, respectively \citep{Lobban18}.}
\label{fig:HST}
\end{figure}       

Figure~\ref{fig:HST} illustrates the wavelength coverage with the OM broad-band UV-filters of the UV emission of Ark\,120 
by comparison with the observed UV spectrum obtained with the Hubble Space Telescope Faint Object Spectrograph (HST/FOS) 
on 1995 July 29 (post-COSTAR with a science aperture of $0\farcs86$ in diameter) by \cite{kuraszkiewicz04}\footnote{The merged calibrated spectrum of Ark\,120 was downloaded 
from http://hea-www.harvard.edu/$\sim$pgreen/HRCULES.html, and smoothed with a four-bin window for better visibility.}. 
The green data are the UVW2 and UVM2 synthetised photometry from the HST/FOS spectrum using the OM filter profiles. 
The OM broad-band UV photometry is little affected by the bright broad emission lines and is a measure of the continuum emission. 
The red and blue data, respectively, are the observed OM UV photometry on 2014 March 22 and on 2013 February 18 \citep{Lobban18}.

We estimated the contribution of the host galaxy to the OM optical photometry
by using the flux variation gradient method proposed by Choloniewski (1981). 
In this method, the combined fluxes of the galaxy and the AGN are obtained 
in two broad-band filters within an aperture centred on the galactic nucleus 
and plotted in a flux-flux diagram. 
The observed flux-flux variation produced by the AGN activity 
follows a linear trend, characterised by a slope determined 
by the host-free AGN continuum in the filter pair, 
and independent of the aperture size as the AGN is spatially unresolved
\cite[e.g.][]{Choloniewski81,Doroshenko08,Winkler92,Haas11}.
Therefore, the AGN colour index in the filter pair is independent of the AGN flux.
In this flux-flux diagram, the galaxy locus is a line going through the origin 
with a slope given by the galaxy colour index in the filter pair.
The intersection of these two lines provides the estimate of the galaxy flux in 
the aperture.

\begin{figure}[t]
\centering
\includegraphics[angle=0,width=\columnwidth]{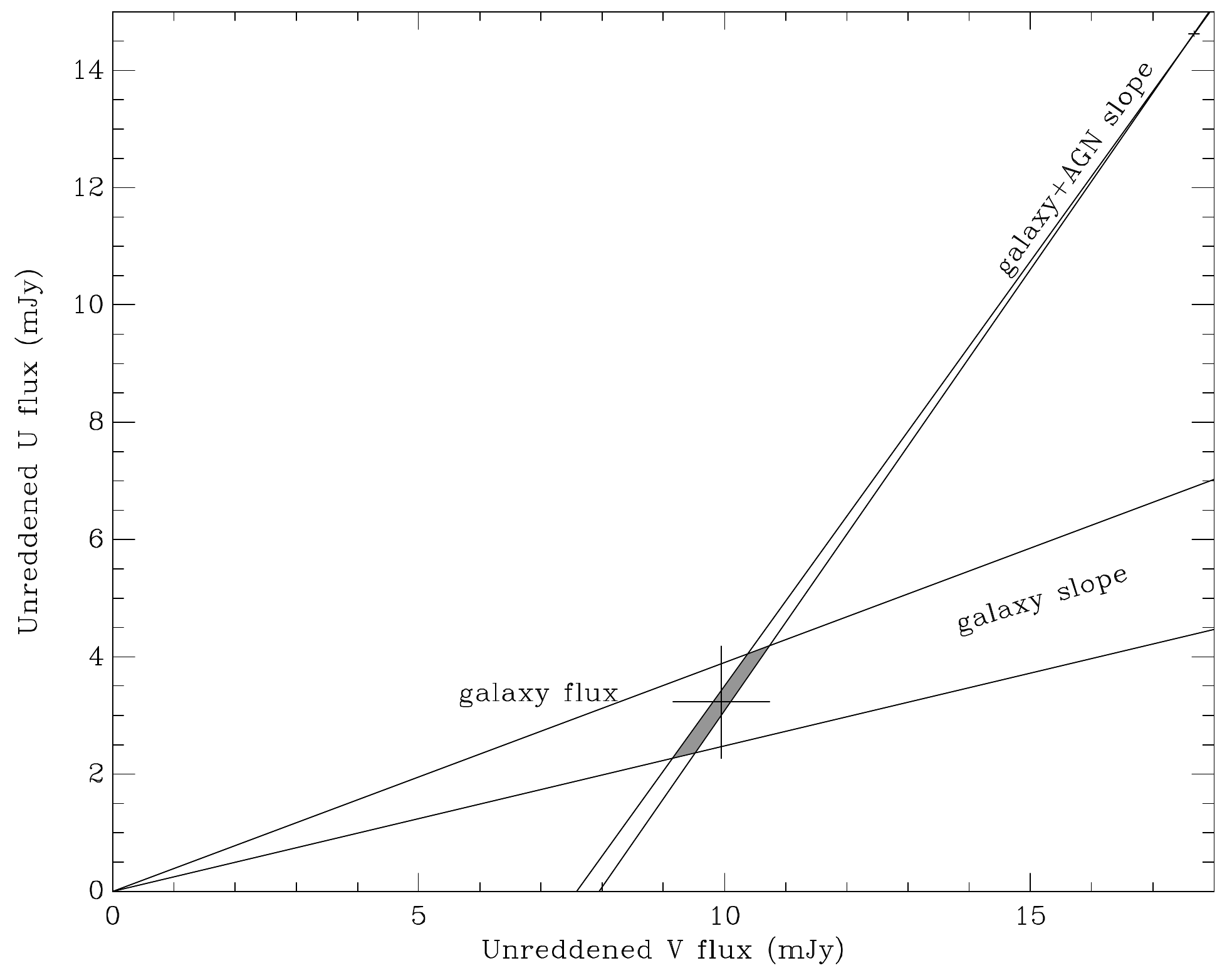}
\caption{$U$ versus $V$ dereddened fluxes of Ark 120, measured in a $6\farcs7$-radius aperture with OM. 
The galaxy and galaxy+AGN slopes are from \cite{Doroshenko08}.}
\label{fig:UV}
\includegraphics[angle=0,width=\columnwidth]{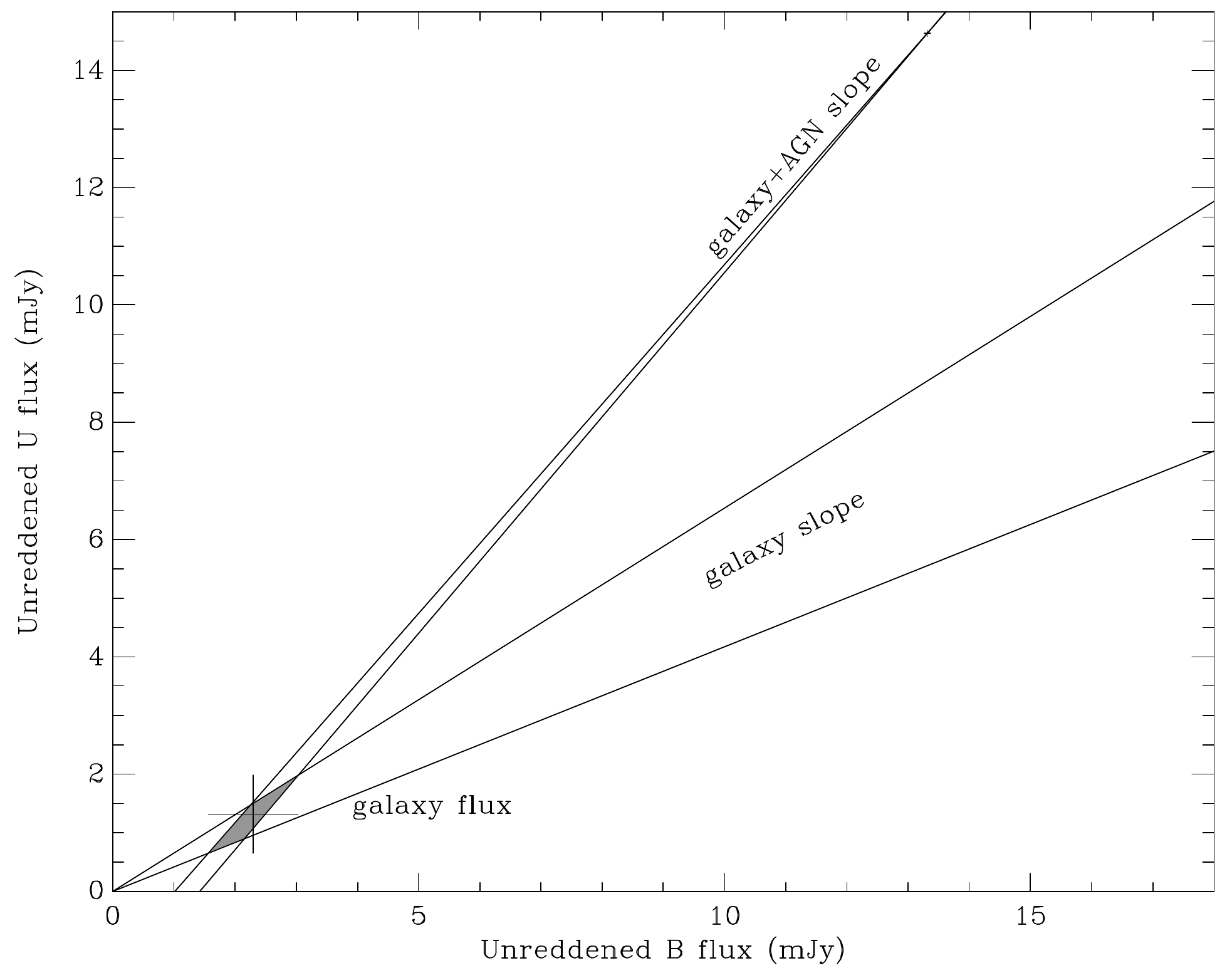}
\caption{$U$ versus $B$ dereddened fluxes of Ark 120, measured in a $6\farcs7$-radius aperture with OM. 
The galaxy and galaxy+AGN slopes are from \cite{Doroshenko08}.}
\label{fig:UB}
\end{figure}

We selected near-simultaneous OM photometry in $V$-$U$ and $U$-$B$ filters obtained 
in subsequent exposures\footnote{Namely, $V$-filter S403 and $U$-filter S007 exposures, 
and $U$-filter S408 and $B$-filter S008 exposures.}.
To convert the OM count rates to fluxes we used the conversion factors obtained from 
observations of standard white dwarf stars\footnote{Namely, $FCFV=2.49$, $FCFB=1.29$, $FCFU=1.94$ in unit 
of $10^{-16}$~erg\,cm$^{-2}$\,\AA$^{-1}$\,count$^{-1}$
(from the header of the calibration file OM$\_$COLORTRANS$\_$0010.CCF).} to be consistent 
with the canned OM response matrices used for the spectral modelling 
in Sect.~3. We dereddened these observed fluxes using $E(B-V)=0.113$ \citep{Schlafly11}  
and the extinction law of \cite{Cardelli89} using $R_\mathrm{V}=3.1$,
leading to $A_\mathrm{V,OM}=0.355$~mag, $A_\mathrm{B,OM}=0.451$~mag, and $A_\mathrm{U,OM}=0.563$~mag.
We obtain: $F_V=17.68 \pm 0.09$~mJy, $F_B=13.32 \pm 0.05$~mJy, and $F_U=14.63 \pm 0.05$~mJy.

We used the AGN and galaxy colour indices obtained by \cite{Doroshenko08} for Ark~120 in 
$7\farcs5$-radius aperture and $7\farcs7$-$13\farcs8$--radius annulus, respectively, 
neglecting possible variations of the galaxy colour indices with distance from the nucleus.
Following \cite{Doroshenko08}, we used $U$-$V$ and $U$-$B$ filter pairs 
to minimise the impact of the colour-index errors on the intersection, 
as the differences between the colour indices of the AGN and galaxy is largest for these filter pairs. 
From the AGN colour indices $U-B=-1.015\pm0.018$ and $B-V=+0.021\pm0.009$ (see Table 3 of \citealt{Doroshenko08})
and after dereddening\footnote{We adopted for the effective wavelengths 5\,500, 4\,330, 3\,650~\AA\ 
for the $V$, $B$, and $U$ Johnson filters, respectively. 
We dereddened of $A_\mathrm{V}=0.350$~mag, $A_\mathrm{B}=0.473$~mag, and $A_\mathrm{U}=0.545$~mag.} 
we computed\footnote{We adopted for the conversion 
from magnitude in the Johnson-Cousin photometric system to Jy, 
the zero magnitude fluxes of 
3\,836.3, 4\,266.7, and 1\,895.8~Jy in $V$, $B$, and U, respectively.} 
the galaxy+AGN flux-flux slopes of 
$1.48\pm0.03$ and $1.21\pm0.02$ for the $UV$ and $UB$ flux-flux plot, respectively. 
From the galaxy colour indices $(U-B)_\mathrm{g}=-0.13\pm0.24$ and $(B-V)_\mathrm{g}=+0.80\pm0.03$ 
\citep{Doroshenko08} and after dereddening, we computed the galaxy flux-flux slopes of  
$0.32\pm0.07$ and $0.54\pm0.12$ for the $UV$ and $UB$ flux-flux plot, respectively. 

Figures~\ref{fig:UV} and \ref{fig:UB} show the $UV$ and $UB$ flux-flux plots, respectively. 
Taking into account the slope uncertainties, we obtain the following dereddened fluxes of the galaxy 
in the central aperture: 
$F_V^\mathrm{g}=9.9 \pm 0.8$~mJy, $F_B^\mathrm{g}=2.3 \pm 0.7$~mJy, 
and $F_U^\mathrm{g}=2.3 \pm 1.2$~mJy (average value of the $UV$ and $UB$ flux-flux plots),
corresponding to 
$56 \pm  4\%$, $17 \pm  5\%$, and $15 \pm  7\%$, respectively, 
of the dereddened flux inside the central aperture.
Since the V data are strongly dominated by the host galaxy emission, 
we did not use them for the SED fitting, and used the B value from the UB flux-flux plot.
Therefore, we subtracted the following estimate of the galaxy contributions inside the central aperture to the observed (reddened) count rates:  
$9.8 \pm  3.1$, and $16.7 \pm  8.6$~counts\,s$^{-1}$
in the $B$, and $U$ filters, respectively. 

For each optical and UV filters, we took the average (without weighting) of count rates 
in the unbinned central image (by contrast to {\tt omichain}, we did not include the redundant 
count rate of the binned central image) with corresponding Gaussian propagated errors.
We subtracted the above contribution of the galaxy only for the optical count rates, as the host contribution is negligible in the UV.

In order to take into account the OM calibration uncertainty of the conversion factor between 
the count rate and the flux, we added quadratically to the statistical error of the count rate a 
representative systematic error of 1.5\%\footnote{http://xmm2.esac.esa.int/docs/documents/CAL-SRN-0346-1-0.pdf}. 
The final OM reddened count rates and associated errors are reported in Table~\ref{tab:OM}. 
We used OM canned response matrices\footnote{Available at 
https://www.cosmos.esa.int/web/xmm-newton/om-response-files} to fit this OM photometry with XSPEC.

\begin{table}[t]
\small
\caption{OM average reddened count rates (expressed in $s^{-1}$) of the AGN and associated errors 
for optical and UV filters used for the SED fitting. 
See text for details on their calculations.}
\begin{tabular}{l@{\hspace{5pt}}c@{\hspace{5pt}}c@{\hspace{5pt}}c@{\hspace{5pt}}c@{\hspace{5pt}}c@{\hspace{5pt}}c@{\hspace{5pt}}}
\hline \hline
Obs.\ date  & B & U & UVW1 & UVM2 & UVW2 \\
(yyyy/mm/dd) &\\
\hline
2014/03/22 & 82.3$\pm$6.2 & 101.7$\pm$6.4  & 60.1$\pm$0.9 & 13.8$\pm$0.2  &  5.49$\pm$0.08\\
2013/02/18 &  $\ldots$ & $\ldots$ & 45.6$\pm$0.7  & 9.5$\pm$0.1 & 3.61$\pm$0.06   \\
\hline
\hline
\end{tabular}
\label{tab:OM}
\flushleft
\end{table}

\subsection{{\sl NuSTAR} data reduction}

{\sl NuSTAR} \citep{Harrison13} observed Ark 120 with its two
co-aligned X-ray telescopes with corresponding focal planes: Focal
Plane Module A (FPMA) and B (FPMB) starting on 2013 February 18 and
2014 March 22 for a total of $\sim$166\,ks and $\sim$131\,ks of elapsed
time, respectively.  The Level 1 data products were processed with the
{\sl NuSTAR} Data Analysis Software (NuSTARDAS) package
(v. 1.6.0). Cleaned event files (level 2 data products) were produced
and calibrated using standard filtering criteria with the
\textsc{nupipeline} task and the calibration files available in the
{\sl NuSTAR} calibration database (CALDB: 20170222). 
Extraction radii for both the source and background spectra were $1.25$ arcmin. After
this process, the net exposure times for the two observations were
about 79 ks (2013) and 65 ks (2014).
The two pairs of {\sl NuSTAR} spectra were binned in order to
over-sample the instrumental resolution by at least a factor of 2.5
and to have a Signal-to-Noise Ratio (SNR) greater than five in
each spectral channel. 
We allowed for cross-calibration uncertainties between the two {\sl NuSTAR} spectra and the simultaneous {\sl XMM-Newton}/pn 
spectrum by including in the fit a cross-normalisation constant -- which are let free to vary --
 corresponding to C$_{\rm NuSTAR\,A}$ and C$_{\rm NuSTAR\,B}$ 
for {\sl NuSTAR} FPMA and FPMB spectra, respectively.

\section{Spectral modelling}\label{sec:models}

\subsection{Galactic hydrogen column density and extinction correction}

 The Galactic Hydrogen column density ($N_{\rm H}$) is assumed to be 9.78$\times$10$^{20}$\,cm$^{-2}$ 
as inferred from the weighted average $N_{\rm H}$ value of the Leiden/Argentine/Bonn Survey of
Galactic \ion{H}{i} \citep{Kalberla05}. 
Since there can be some additional contribution associated with
molecular hydrogen \citep{Willingale13}, we allowed the value of
Galactic $N_{\rm H}$ to slightly vary.  
However, we did not allow for any intrinsic absorption in the rest frame of 
Ark\,120, since, as found in \cite{Reeves16} from the 2014 deep RGS spectrum, none
is present. 
We used the X-ray absorption model {\sc tbnew (v2.3.2)} from
\cite{Wilms00}, assuming throughout their ISM elemental abundances and
the cross-sections from \cite{Verner96}. 

The {\sc redden} component allows us to take into account the
 IR-optical-UV extinction from our Galaxy \citep{Cardelli89}. 
 The extinction at V is A(V)=$E$(B$-$V)$\times$~$R_{\rm V}$, with the standard value 
 of $R_{\rm V}$ being 3.1 for the Milky Way. 
We fixed $E$(B$-$V) to 0.113 that corresponds to 
Galactic Extinction from the \cite{Schlafly11} recalibration 
of the \cite{Schlegel98} infrared-based dust map. 
As mentioned previously,  Ark\,120 does not show any intrinsic reddening
 in its infrared-optical continuum \citep{Ward87,Vasudevan09}. 

\subsection{Disc-Comptonisation modelling: optxconv}\label{sec:optxagnf/optxconv}

The broad X-ray spectrum of Ark\,120 is dominated by warm and hot Comptonisation in both March 2014 \citep[][see their Fig.\ 9]{Porquet18}, 
and February 2013 observations \citep{Matt14}, 
though a mild relativistic reflection contribution is observed beyond tens of $R_{\rm g}$ in 2014.  
As described in the introduction, the baseline model used in this work, 
{\sc optxagnf} \citep{Done12}, allows us to infer the global energetics of the flow \citep{Davis11}, 
and then the SMBH spin. 
Since the {\sc optxagnf} model  
includes the colour temperature corrected black body of the outer accretion
disc, we were able to use the
corresponding OM data (see section~\ref{sec:OM}). 

\subsubsection{Description of the optxagnf model parameters}

The {\sc optxagnf} model is characterised by the
following parameters: \\
$-$ The SMBH mass ($M_{\rm BH}$) in solar masses. \\
$-$ The co-moving distance (D) in Mpc.\\
$-$ The log($L_{\rm bol}/L_{\rm Edd}$) ratio, which is equal to log($\dot{M}$/$\dot{M}_{\rm Edd}$), where $\dot{M}$ is the absolute accretion rate (see Eq.~\ref{eq:Lband})  and $\dot{M}_{\rm Edd}$ is the Eddington accretion rate.\\
$-$ The dimensionless BH spin ($a$): 0$\leq$ $a$ $\leq$ 0.998.\\ 	
$-$ The coronal radius ($R_{\rm corona}$) in units of $R_{\rm g}$ where the transition from a colour temperature corrected black body emission to a Comptonised
spectrum occurs (the latter extending down to $R_{\rm ISCO}$). \\
$-$ The log of the outer radius of the disc in units of $R_{\rm g}$:
here, we used the option allowing to fix it to the self-gravity radius as calculated from \cite{Laor89}. 
However, fixing it to a specific value, such as for example five, has only a marginal impact of the fit results.\\
$-$ The electron temperature ($kT_{\rm e}$) of the warm Comptonisation component in
keV.\\
$-$ The optical depth ($\tau$) of the warm Comptonisation component. \\
$-$ The spectral index ($\Gamma$) of the hot Comptonisation component
(power-law shape) which has a (fixed) temperature set internally at 100 keV (based on the {\sc nthcomp} 
model; \citealt{Zdziarski96,Zycki99}).\\
$-$ The fraction ($f_{\rm pl}$) of the power below $R_{\rm corona}$ which is emitted 
in the hot Comptonisation component. 

\subsubsection{Taking into account inclination effects}

The {\sc optxagnf} model is by default calculated for a disc inclination angle ($\theta$) of
60$^{\circ}$ (that is, for a normalisation value tied to unity). Therefore in
order to take into account inclination effects on the {\sc
  optxagnf} component emission, we linked its normalisation to the
disc inclination angle of the relativistic reflection component ({\sc
  reflcomp}, see section~\ref{sec:reflection}) using a {\sc cos($\theta$)/\sc cos$(60^\circ)$} relationship. 

\subsubsection{Taking into account relativistic effects: the optxconv model}

The {\sc optxagnf} model does not
include any relativistic effects on the propagation of light from the disc to the observer, but the
combination of Doppler and gravitational shifts may impact the results.
We included these relativistic effects using the  {\sc optxconv} model detailed in \cite{Done13}, 
which effectively convolves the broad-band emission from {\sc optxagnf} with the relativistic blurring
calculated by the {\sc relconv} convolution model \citep[v0.4c;][]{Dauser10}. 
The {\sc relconv} model is characterised by the following parameters:\\
$-$ the radius ($R_{\rm br}$ expressed in
$R_{\rm g}$) where the broken power-law emissivity index changes from $q_1$ (for $R$$<$$R_{\rm br}$) to $q_2$ (for $R$$>$$R_{\rm br}$).
 Throughout this work, $q_1$ and $q_2$ are tied together and fixed to the typical value of 3.0.\\
$-$ the dimensionless BH spin ($a$);\\
$-$ the disc inclination angle ($\theta$, expressed in degrees);\\
$-$ the inner and outer radii of the disc where the relativistic
reflection is observed: $R_{\rm in}$ and $R_{\rm out}$ (expressed in $R_{\rm g}$),
respectively;\\
$-$ The limb-darkening/-brightening laws (0, 1 and 2 correspond to
isotropic, darkening and brightening law, respectively). Here, we assumed 
an isotropic value but we checked that this has a negligible impact on the fit
results. \\

The soft excess (warm Comptonisation) and the hard energy tail (hot 
Comptonisation) were convolved with 
the {\sc relconv} relativistic kernel between $R_{\rm ISCO}$ and $R_{\rm corona}$ where the coronal emission occurs.   
Outside $R_{\rm corona}$, that is, the colour temperature corrected black body spectrum (here called the outer disc emission)
at each radius could be in principle convolved with {\sc relconv} at that radius, and  integrated from $R_{\rm corona}$ to $R_{\rm out}$. 
However, this would be extremely time-consuming to calculate. 
Therefore,  we used the fact that most of the outer disc emission arises from radii less than twice that of its innermost radius,
that is, from  $R_{\rm corona}$ to 2\,$R_{\rm corona}$ (see \citealt{Done13}). Indeed, at much larger disc radii far from the black hole, the 
relativistic correction to {\sc optxagnf} are largely negligible. 

	The {\sc optxconv} model can be summarised as:\\
 {\sc relconv}[$R_{\rm in}$=$R_{\rm corona}$;$R_{\rm out}$=2$R_{\rm
   corona}$]$\otimes${\sc optxagnf}(outer disc) + {\sc relconv}[$R_{\rm in}$=$R_{\rm
   ISCO}$;$R_{\rm out}$=$R_{\rm corona}$]$\otimes${\sc optxagnf}(warm and hot
 Comptonisation).

\subsection{Relativistic reflection modelling: the {\sc reflcomp} model}\label{sec:reflection}

The 2013 and 2014 X-ray broad-band spectra are dominated by warm and hot Comptonisation \citep{Matt14,Porquet18}, nonetheless in 2014 an additionnal mild relativistic reflection component is still required \citep{Nardini16,Porquet18}. Its low degree of relativistic smearing indicates  that this emission only arises from a few tens of gravitational radii rather than closer to the ISCO. However, this reflection component becomes non negligible above about 10\,keV and must be included in the spectral fits for an accurate determination of the bolometric luminosity (Eq.~\ref{eq:Lbol}). Even if no significant relativistic reflection component is reported for the 2013 February 18 observation \citep{Matt14}, we included one
for self-consistency, and checked for any possible contribution.  
 
The relativistic reflection component, hereafter called {\sc reflcomp}, was calculated 
using the following model: {\sc relconv$\otimes$xilconv$\otimes$optxagnf}.

The {\sc xilconv} convolution model \citep{Done06,Kolehmainen11} 
combines an ionised disc table model from the {\sc xillver}  model
\citep{Garcia10} with the \cite{Magdziarz95} Compton reflection code. 
 Therefore, it allows us to use the optxagnf (Comptonisation) emission as the incident spectrum 
for the reflection component, that is, xilconv$\otimes$optxagnf. 
In order to account for relativistic effects, we also convolved it with
the {\sc relconv} model.

The {\sc xilconv} model is characterised by the following parameters: \\ 
$-$ the reflection fraction: $\mathcal{R}$; \\
$-$ the iron abundance relative to the solar value (\citealt{Grevesse98}):
$A_{\rm Fe}$. Here, we fixed it to unity, but if let free to vary 
 it has only a marginal impact on the fit results; \\
$-$ the ionisation parameter (erg\,cm\,s$^{-1}$, in log units) at the
surface of the disc  
(that is, the ratio of the X-ray flux to the gas density): log\,$\xi$; \\ 
$-$ the high-energy cut-off: $E_{\rm cut}$ (expressed in keV).
 Here, its value was fixed to 300 keV to be consistent with the temperature internally set at 100\,keV  in {\sc optxagnf} (section~\ref{sec:optxagnf/optxconv}) 
 -- which is taken as the incident spectrum for the relativistic reflection model --
  as for a thermal distribution of electrons $E_{\rm cut}$ $\sim$ 3\,$kT$ \citep[e.g.][]{Fabian15}.\\
The parameters of the {\sc relconv} and {\sc optxagnf} models have been already described in section~\ref{sec:optxagnf/optxconv}.  

In this work, we assumed that the warm optically-thick corona has
a full coverage. 
Therefore, unless otherwise mentioned we fixed $R_{\rm in}$ (inner
radius of the observed relativistic reflection) to $R_{\rm corona}$
(corona radius)  meaning that below $R_{\rm corona}$ any reflection is hidden to the
 observer by the warm optically-thick corona.

\subsection{Modelling of the Fe\,K complex components from the broad line region (BLR)}

In \cite{Nardini16}, from modelling the 2014 {\sl Chandra} HETG observation of Ark 120,
  we resolved the core of the Fe K$\alpha$ line, where its velocity width (FWHM$\sim$ 4700 km\,s$^{-1}$)
  was found to be consistent with the broad H$\beta$ line in the optical spectrum.
  Thus, the narrow neutral core of the Fe K$\alpha$ emission is assumed to be associated
  with the optical BLR rather than the torus, which may also be the case for any ionised emission from Fe XXVI Ly$\alpha$.
Therefore, throughout this work  
we took into account the contribution from the BLR emission 
to the Fe\,K complex using three Gaussian lines (as in \citealt{Porquet18}):
 the Fe\,K$\alpha_{\rm BLR}$ ($E$ fixed at 6.40\,keV) plus its
 associated Fe\,K$\beta_{\rm BLR}$ 
 line ($E$ fixed at 7.05\,keV), and the H-like iron line ($E$ fixed at
 6.97\,keV).  
 The normalisation of Fe\,K$\beta_{\rm BLR}$ was set to 0.135 times
 that of Fe\,K$\alpha_{\rm BLR}$ \citep{Palmeri03}.   
The widths of these three lines were fixed to the value inferred 
for the Fe\,K$\alpha$ narrow core, that is, 43\,eV as
determined from the simultaneous {\sl Chandra}/HETG spectrum \cite{Nardini16}. 
These three BLR emission lines are called hereafter `3$\times$
{\sc zgaussian}(BLR)'.

\section{Spectral analysis}\label{sec:analysis}

The {\sc xspec v12.9.1}p software package \citep{Arnaud96}  
was used for spectral analysis. 
We used throughout the $\chi^{2}$ minimisation, quoting confidence levels of  
90 percent for one interesting parameter ($\Delta\chi^{2}$=2.71). 
Unless stated otherwise, we assumed a reverberation SMBH mass value of
1.50$\times$10$^{8}$\,M$_{\odot}$ \citep{Peterson04}. 
We adopted a distance\footnote{This distance value is calculated via the NED Cosmology Calculator 
\citep{Wright06}, assuming a flat Universe: http://www.astro.ucla.edu/$\sim$wright/CosmoCalc.html.} 
of 143.5\,Mpc inferred from the cosmological constants from
 \cite{Planck16}, that is, $H_{\rm
  0}$=67.8\,km\,s$^{-1}$\,Mpc$^{-1}$, $\Omega_{m}=0.308$, and $\Omega_{\Lambda_{\rm 0}}=0.692$.  
All spectra are displayed in the AGN rest-frame.   
Hereafter, the energy range over which the models are
calculated has been extended up to 500\,keV and down to 1\,eV (B filter band).

Throughout this work, we used the following model 
(where the components have been described in section~\ref{sec:models}) 
simultaneously to the SED of both the 2014 March 22 
and the 2013 February 18 {\sl XMM-Newton} (OM + pn) and {\sl NuSTAR} observations: \\
{\sc redden$\times$tbnew$\times$\{optxconv+ reflcomp +3$\times$zgaussian(BLR)\}}.\\

Since the Galactic column density 
($N_{\rm H}$), the disc inclination angle ($\theta$), and the BH spin (a) 
are not supposed to vary in a year time-scale, they were tied between both datasets. 
Moreover, the disc ionisation parameter was also tied between both datasets 
since its value is similar for 2013 and 2014 observations. 

\begin{figure}[t!]
\begin{tabular}{c}
\includegraphics[width=0.9\columnwidth,angle=0]{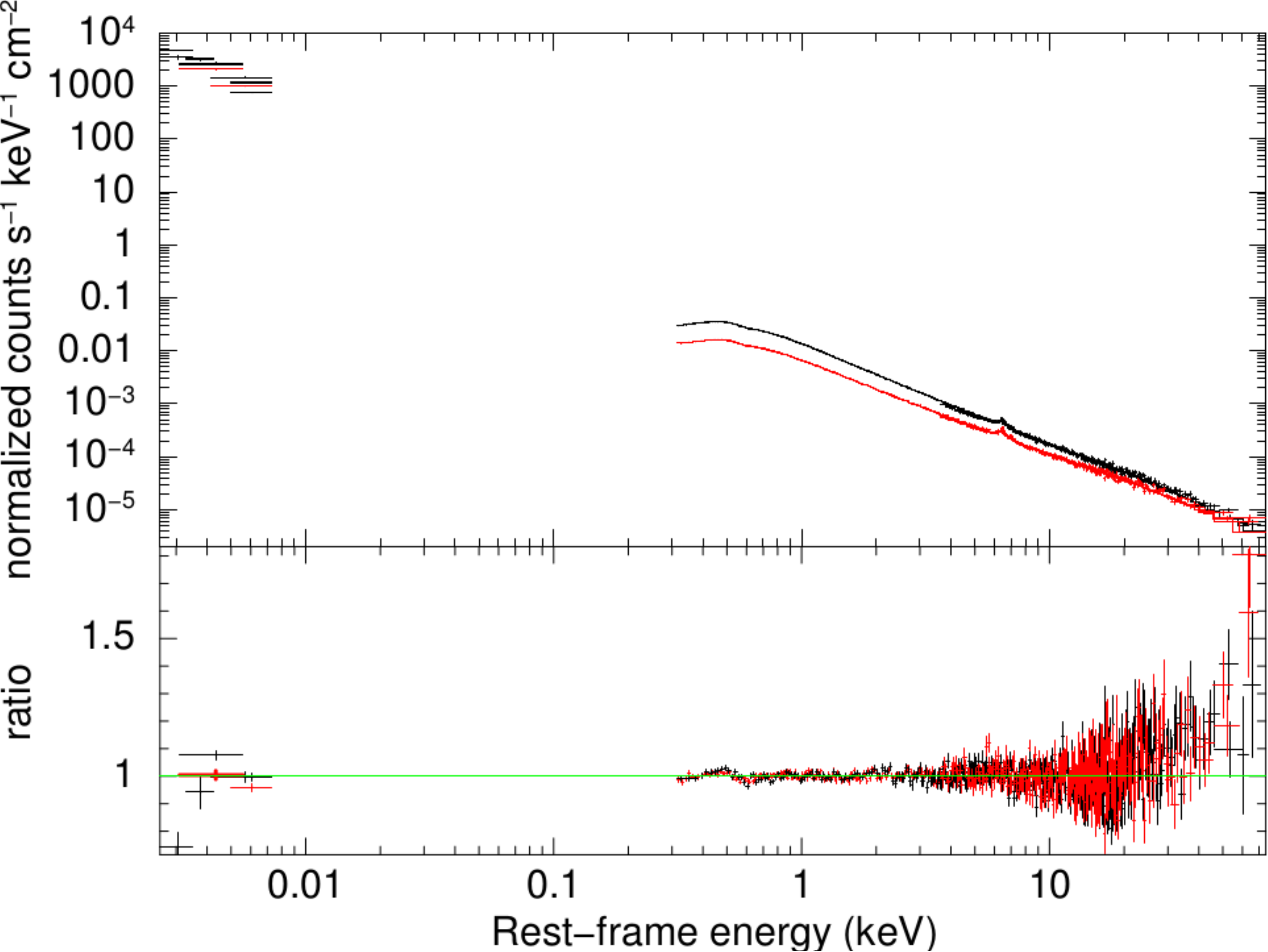} \\
\hspace*{0.19cm}\includegraphics[width=0.86\columnwidth,angle=0]{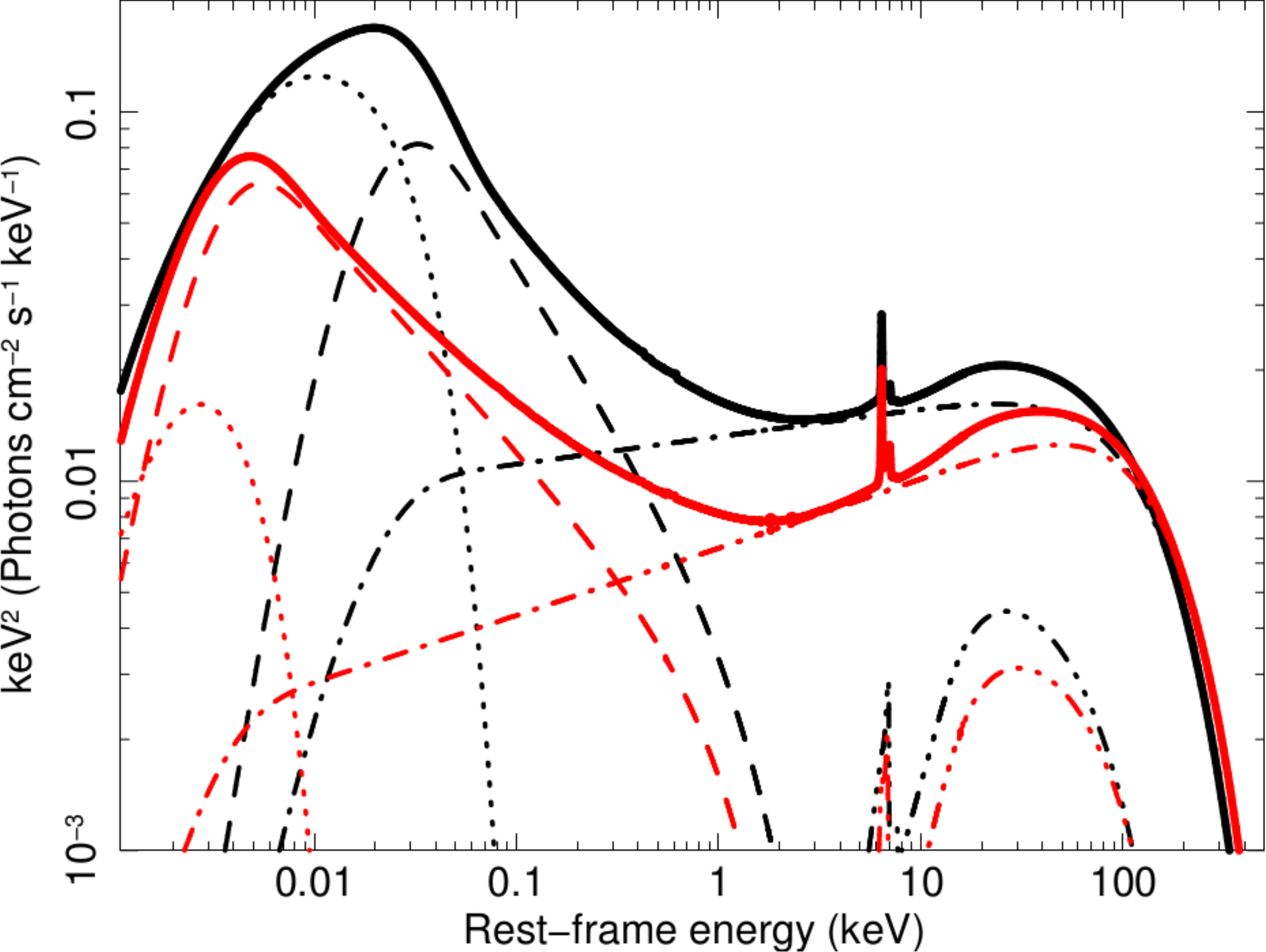}
\end{tabular}
\caption{Simultaneous fit of the Ark\,120 SED ({\sl XMM-Newton/OM/pn} and {\sl NuSTAR}) spectra of Ark\,120 (AGN rest-frame) obtained
  on 2014 March 22 and on 2013 February 18. The 2014 and 2013 observations are displayed in black and red, respectively. 
  Top panel: spectra and data/model. Bottom panel: intrinsic (that is, corrected for reddening and Galactic absorption) SED and the corresponding model components. Continuous curves: total SED. Dotted curves: outer disc emission. Dashed curved: warm optically-thick Comptonisation component (soft excess). Dotted-dashed curves: hot Comptonisation component (hard energy tail). 3-dotted-dashed curves: relativistic reflection component.
For clarity purposes, the three BLR Gaussian line components are not displayed.}
\label{fig:2013-2014-modelB}
\end{figure}

We find a good fit statistic ($\chi^{2}$/d.o.f.=3824.4/3517, $\chi^{2}_{\rm red}$=1.09), 
 the best fit parameters of this model are listed in Table~\ref{tab:optxconv} (left-hand column). 
However, some deviations are found in the hardest energy part of both X-ray spectra 
(Fig.~\ref{fig:2013-2014-modelB}, top panel). 
In {\sc optxagnf} the temperature of the hot component is not a free parameter since internally fixed to 100\,keV 
(see section~\ref{sec:optxagnf/optxconv}). 
The best-fit temperature may vary from this value during these 2014 and 2013 observations. 
Indeed, \cite{Marinucci19}, using a {\sc nthcomp} component (as used in {\sc optxagnf} and {\sc optxconv} models) 
for the hot corona find kT$_{\rm e}$\,$\geq$\,40\,keV for the 2013 dataset and 155$^{+350}_{-55}$\,keV for the 2014 dataset.  
However, these deviations have a very marginal impact on the determination of the bolometric luminosity and then on the inferred spin value found in this work. 
There is also a deviation of the B band flux for 2014 that may be due to an 
 overestimation of the true galaxy host contribution at this wavelength.   

For the 2014 observation, we infer temperature and optical depth 
values for the warm Comptonisation (producing part of the soft excess) that are similar to
those found in \cite{Porquet18}, where a simplified modelling with the {\sc comptt} model has been used. 
It is worth pointing out, that during this 2014 observation,
both the flux and photon index correspond to the `high-flux spectrum'
found for Ark\,120 from a {\sl Swift} monitoring \citep{Gliozzi17,Lobban18}, 
 as illustrated in Fig.~\ref{fig:swift}. 
The optically-thick corona
extension is rather moderate with $R_{\rm corona}$ of about 14\,$R_{\rm g}$. 
We infer an accretion rate consistent with the one inferred from the $\sim$ six-month 
{\sl Swift} monitoring data of Ark\,120 \citep{Buisson17}.
 
\begin{figure}[!ht]
\begin{tabular}{c}
\includegraphics[width=0.9\columnwidth,angle=0]{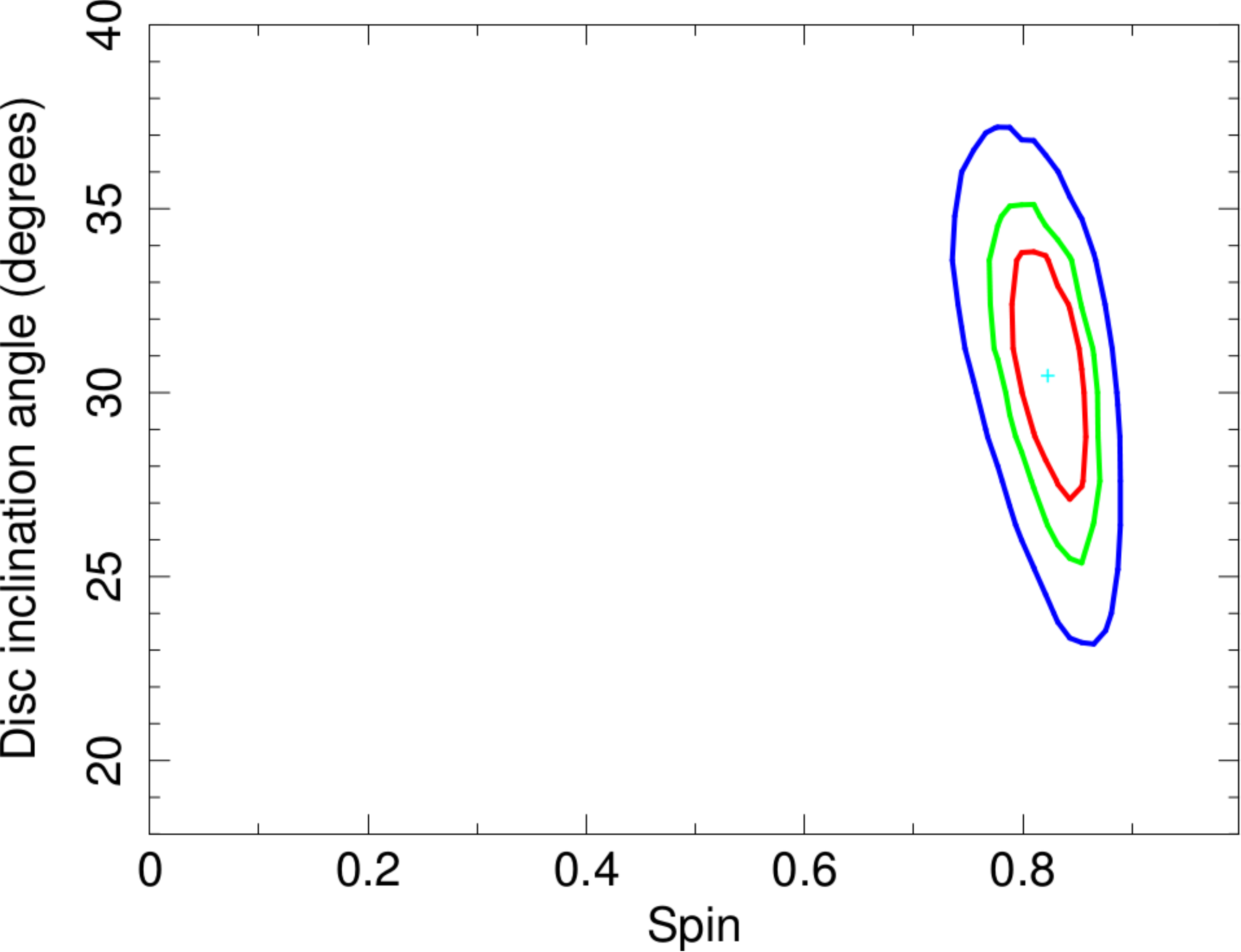} \\
\includegraphics[width=0.9\columnwidth,angle=0]{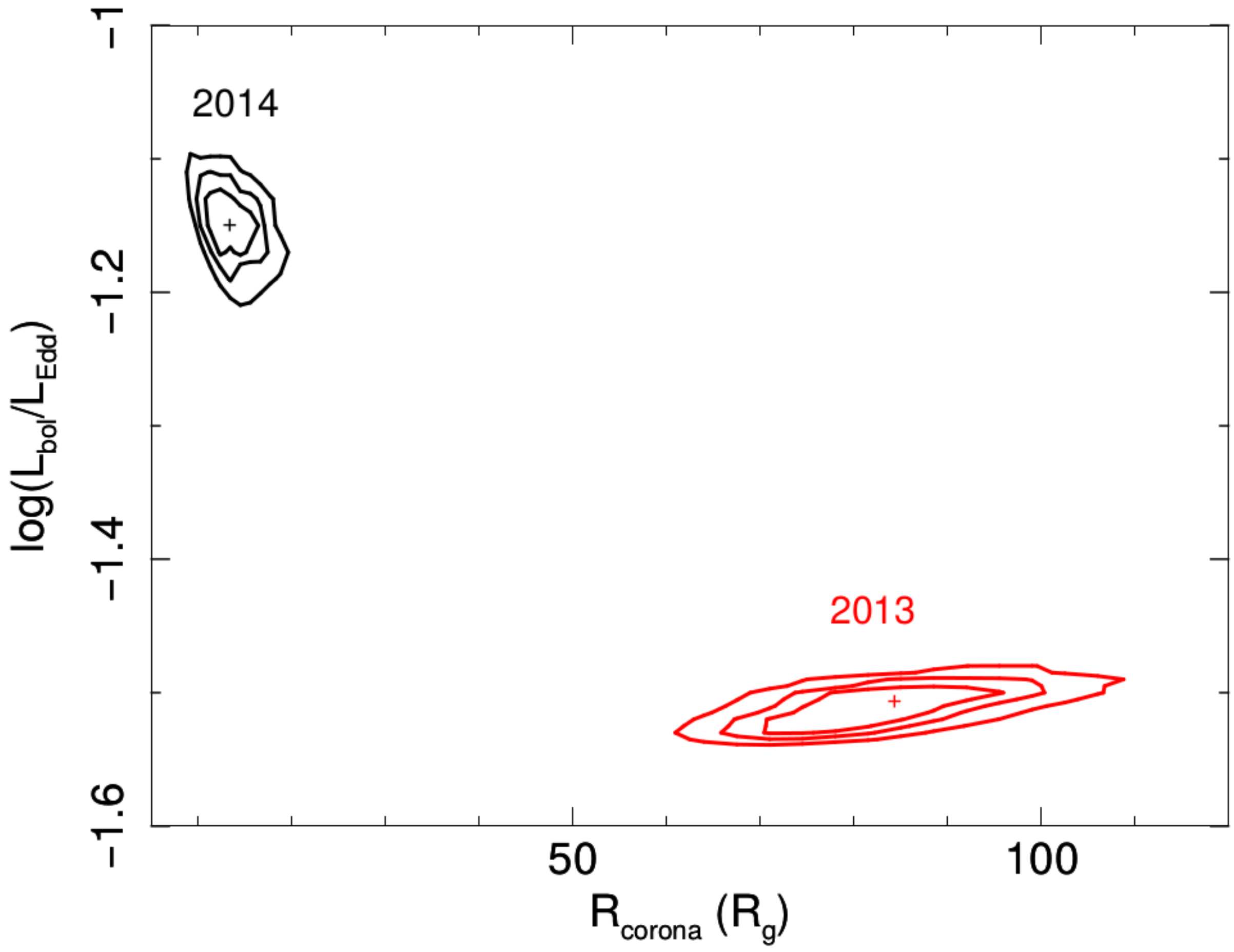} \\ 
\end{tabular}
\caption{2D contour plots (at the 68\%, 90\% and 99\% confidence levels) 
from the simultaneous fit of the 2013 and 2014 observations. 
Top panel: disc inclination angle versus spin.
Bottom panel: log($L_{\rm bol}/L_{\rm Edd}$) versus $R_{\rm corona}$ (expressed in $R_{\rm g}$) found for the 2014 observation (high-flux state, in black) and for the 2013 observation (low-flux state, in red).
}
\label{fig:2013-2014-contours}
\end{figure}

 We find the overall observed luminosity requires a spin value of
0.83$^{+0.05}_{-0.03}$ which is remarkably well constrained. 
Since relativistic reflection component ({\sc reflcomp}) is only observed beyond about 
14\,$R_{\rm g}$ in 2014, the spin constraint is driven by the {\sc optxconv} SED modelling, however the disc inclination -- that is an important variable to determine the accretion rate (see Eq.~\ref{eq:Lband}) -- is well constrained thanks to the  
mildly relativistic reflection component present during the 2014 observation. 
Therefore, this shows that in order to obtain via this method a good constraint on the spin value,  the presence of a relativistic reflection component is needed in order to measure the disc inclination.   
Hence, if we consider the 2013 data in isolation, we are unable to constrain the spin value, due to the lack of any disc relativistic reflection component (also see \citealt{Matt14}).   
We notice that the inferred value of the accretion disc inclination ($\theta$\,$\sim$30\,$^{\circ}$) is similar to that of the host galaxy \citep[that is, 26$^{\circ}$;][]{Nordgren95}. 
Figure~\ref{fig:2013-2014-contours}   
displays the 2D contour plots of the spin versus the disc inclination
angle (top panel).

We find that the temperature of the warm optically-thick corona has
slightly increased with statistical significance of about 1.8$\sigma$ between 2013 and 2014 
 (see Table~\ref{tab:optxconv}, column 1), 
corresponding to unabsorbed 0.3--2\,keV fluxes of 2.58\,$\times$\,10$^{-11}$\,erg\,cm$^{-2}$\,s$^{-1}$ and 5.42\,$\times$\,10$^{-11}$\,erg\,cm$^{-2}$\,s$^{-1}$, respectively. 
Moreover, the photon index of the hot optically-thin corona has softened (by about 3.9$\sigma$) 
between the 2013 ($\Gamma$=1.82$^{+0.02}_{-0.01}$) and 2014 ($\Gamma$=1.93$\pm$0.02)  observations, corresponding to 2--79\,keV flux of 7.28$\times$10$^{-11}$\,erg\,cm$^{-2}$\,s$^{-1}$ and 1.03$\times$10$^{-10}$\,erg\,cm$^{-2}$\,s$^{-1}$, respectively. This again confirms previous results \citep{Matt14,Gliozzi17,Lobban18} where a
`softer when brighter' behaviour for Ark\,120 was reported. 

\begin{table*}[!ht]
\caption{Simultaneous SED (from optical to hard X-rays) fitting of both 2014 March 22 and 2013 
  February 18 observations with the model {\sc redden$\times$tbnew$\times$\{optxconv+ reflcomp +3$\times$zgaussian(BLR)\}} described in section~\ref{sec:models}. We allow for cross-calibration uncertainties between the two {\sl NuSTAR} spectra and the simultaneous {\sl XMM-Newton}/pn spectrum by including in the fit a cross-normalisation constant corresponding to $C_{\rm NuSTAR\,A}$ and $C_{\rm NuSTAR\,B}$ for {\sl NuSTAR} FPMA and FPMB spectra, respectively. The fit with the default
  values for $E$(B$-$V), the AGN distance,  and $M_{\rm BH}$ are reported in
  column\,1. The other columns report the fit results when we vary one
  assumption from the fixed values (marked in bold) compared to
  the default values of $E$(B$-$V), the AGN distance and $M_{\rm BH}$. (f) means that the parameter value is fixed.}             
\label{table:1}    
\centering                          
\begin{tabular}{l| c c c c c }        
\hline\hline                 
$E$(B$-$V) & 0.113 (f) & {\bf 0.128 (f)}   & 0.113 (f)  & 0.113 (f)   &  0.113 (f)   \\
$D$ (Mpc) & 143.5 (f) & 143.5 (f)   & {\bf 137.2 (f)}    & 143.5 (f) &   143.5 (f)  \\
$M_{\rm BH}$ ($\times$10$^{8}$\,$M_{\odot}$) & 1.50$^{(a)}$ (f) & 1.50$^{(a)}$ (f) & 1.50$^{(a)}$ (f)&
{\bf 1.17$^{(b)}$} (f) & {\bf 1.71$^{(c)}$} (f)\\
\hline
$N_{\rm H}$ ($\times$10$^{20}$\,cm${-2}$) & 9.92$^{+0.15}_{-0.18}$ & 10.03$^{+0.12}_{-0.15}$ &  9.93$^{+0.13}_{-0.20}$ & 9.96$^{+0.11}_{-0.23}$ & 9.95$^{+0.12}_{-0.21}$  \\ 
$a$ & 0.83$^{+0.05}_{-0.03}$ & 0.79$\pm$0.05 & 0.85$^{+0.05}_{-0.03}$ & 0.68$^{+0.05}_{-0.06}$ & 0.89$^{+0.04}_{-0.02}$\\
$\theta$ (degrees)  &  30.3$^{+3.6}_{-13.9}$ & 30.8$^{+3.7}_{-11.4}$   & 30.8$^{+3.2}_{-11.7}$    &  30.5$^{+3.6}_{-10.7}$&  30.8$^{+3.2}_{-11.4}$\\
\hline                                  
                          &   \multicolumn{5}{c}{2014 March 22} \\
\hline                                  
log($L_{\rm bol}/L_{\rm Edd}$)$^{(d)}$  & $-$1.15$\pm$0.03 & $-$1.12$^{+0.02}_{-0.03}$ 
&$-$1.18$^{+0.03}_{-0.04}$   & $-$1.04$\pm$0.03 & $-$1.19$^{+0.02}_{-0.04}$ \\
$kT_{\rm e}$ (keV)&  0.49$^{+0.10}_{-0.05}$ & 0.51$^{+0.08}_{-0.07}$ &
0.49$^{+0.10}_{-0.04}$ & 0.48$^{+0.07}_{-0.05}$ & 0.52$^{+0.08}_{-0.06}$ \\
 $\tau$ & 9.1$^{+0.5}_{-1.0}$  & 8.8$^{+1.5}_{-0.7}$ & 9.1$^{+0.9}_{-1.0}$ & 9.1$^{+1.1}_{-0.7}$ & 8.8$^{+1.2}_{-0.7}$ \\
 $\Gamma$ & 1.93$\pm$0.02 & 1.93$^{+0.01}_{-0.02}$ & 1.93$\pm$0.02 &
 1.93$\pm$0.02 & 1.93$\pm$0.02 \\
 $f_{\rm pl}$  & 0.41$^{+0.06}_{-0.02}$ & 0.36$^{+0.04}_{-0.03}$ &
 0.41$^{+0.05}_{-0.04}$ & 0.42$^{+0.07}_{-0.03}$ & 0.40$^{+0.05}_{-0.03}$ \\
$R_{\rm corona}$ ($R_{\rm g}$) & 13.8$\pm$3.2 & 16.4$^{+3.7}_{-3.5}$
& 12.8$^{+3.4}_{-2.4}$ & 16.5$^{+3.9}_{-4.3}$& 12.1$^{+2.7}_{-3.3}$\\
$\mathcal{R}$ & 0.23$^{+0.05}_{-0.04}$ & 0.23$^{+0.03}_{-0.04}$  &0.23$\pm$0.04  
&0.23$^{+0.05}_{-0.04}$ &  0.23$\pm$0.04   \\
 log\,$\xi$ & $\leq$0.1 & $\leq$0.1  & $\leq$0.1 & $\leq$0.1 & $\leq$0.1 \\
C$_{\rm NuSTAR A}$ & 1.027$^{+0.009}_{-0.008}$ & 1.027$^{+0.009}_{-0.008}$ &
1.027$\pm$0.009  & 1.028$\pm$0.009  & 1.027$\pm$0.009 \\
C$_{\rm NuSTAR B}$ & 1.070$^{+0.007}_{-0.009}$ & 1.070$\pm$0.009  &
1.070$\pm$0.009 &  1.070$\pm$0.009  & 1.070$\pm$0.009 \\
\hline                                  
                          &   \multicolumn{5}{c}{2013 February 18} \\
\hline  
log($L_{\rm bol}/L_{\rm Edd}$)$^{(d)}$  & $-$1.51$^{+0.01}_{-0.03}$ & $-$1.48$^{+0.01}_{-0.03}$
&$-$1.54$^{+0.01}_{-0.03}$   & $-$1.40$^{+0.01}_{-0.02}$ & $-$1.56$^{+0.01}_{-0.03}$\\
$kT_{\rm e}$ (keV)&  0.35$\pm$0.02 & 0.35$\pm$0.02 &
0.35$^{+0.03}_{-0.01}$ & 0.34$^{+0.03}_{-0.02}$ &  0.35$^{+0.04}_{-0.02}$\\
 $\tau$ & 12.1$^{+0.7}_{-0.5}$ & 11.9$^{+0.4}_{-0.3}$  & 12.2$\pm$0.6 &
 12.3$\pm$0.6  & 12.2$\pm$0.6\\
 $\Gamma$ & 1.82$^{+0.02}_{-0.01}$ & 1.82$^{+0.01}_{-0.02}$ & 1.82$\pm$0.01 &
 1.82$\pm$0.02 & 1.82$\pm$0.02   \\
 $f_{\rm pl}$ & 0.32$\pm$0.01 & 0.30$\pm$0.01 & 0.32$\pm$0.01
 & 0.32$\pm$0.01 & 0.32$\pm$0.01 \\
$R_{\rm corona}$ ($R_{\rm g}$) & 84.7$^{+12.5}_{-9.7}$ & 77.6$^{+9.6}_{-8.0}$ 
& 80.9$^{+11.2}_{-9.8}$ & 110.0$^{+18.2}_{-10.7}$ & 73.0$^{+10.5}_{-8.7}$   \\
$\mathcal{R}$ & 0.21$^{+0.05}_{-0.02}$ & 0.21$\pm$0.04 & 0.22$^{+0.05}_{-0.04}$ & 0.21$^{+0.05}_{-0.03}$  & 0.22$^{+0.02}_{-0.05}$   \\
C$_{\rm NuSTAR\,A}$ & 1.057$^{+0.011}_{-0.010}$  & 1.057$\pm$0.011 &
1.057$\pm$0.011  & 1.057$^{+0.011}_{-0.010}$  & 1.057$\pm$0.011 \\
C$_{\rm NuSTAR\,B}$ & 1.075$\pm$0.011 & 1.075$\pm$0.011  &
1.075$\pm$0.011 &  1.075$\pm$0.011 &  1.075$\pm$0.011  \\
\hline \hline
 $\chi^{2}$/d.o.f. & 3824.4/3517 & 3808.7/3517 & 3825.0/3517   &
 3820.6/3517 &  3826.9/3517  \\
 $\chi^{2}_{\rm red}$&1.09  &    1.08 & 1.09  &  1.09 & 1.09 \\
\hline    \hline                  
\end{tabular}
\tablefoot{(a) BH mass from reverberation mapping \citep{Peterson04}
using the calibration of the M--$\sigma_{\star}$ relation for AGNs (mean virial
factor: $\langle f \rangle$=5.5$\pm$1.8) from \cite{Onken04}. (b) BH mass
from reverberation mapping \citep{Peterson04} 
using the calibration of the M--$\sigma_{\star}$ relation for AGNs based on
high-luminosity quasars hosts  (mean virial
factor: $\langle f \rangle$=4.31$\pm$1.05) from \cite{Grier13}
\citep{Bentz15}. (c) BH mass from reverberation mapping \citep{Peterson04}
using the calibration of the M--$\sigma_{\star}$ relation for classical bulge galaxies (mean virial
factor: $\langle f \rangle$=6.3$\pm$1.5) from \cite{Ho14}.
 See section~\ref{sec:mass} for a detailed explanation.   
(d) The log($L_{\rm bol}/L_{\rm Edd}$) ratio is equal to log($\dot{M}$/$\dot{M}_{\rm Edd}$), where $\dot{M}$ is the absolute accretion rate (see Eq.~\ref{eq:Lband})  and $\dot{M}_{\rm Edd}$ is the Eddington accretion rate.
 }
\label{tab:optxconv}
\end{table*}

This disc-corona model implies that the radius of the (warm and hot) corona 
($R_{\rm corona}$) has  
significantly decreased (with a statistical significance of about 5.5$\sigma$) between the 2013 and 2014 observations, 
from 85$^{+12.5}_{-9.7}$ to 14$\pm$3\,$R_{\rm g}$. 
Figure~\ref{fig:2013-2014-contours} (bottom panel)   
displays the 2D contour plots of $L_{\rm bol}/L_{\rm Edd}$ versus $R_{\rm corona}$ (expressed in $R_{\rm g}$) found for the 2014 observation (high-flux state, in black) and for the 2013 observation (low-flux state, in red). 
Consequently, the decreasing inner radius of the observed relativistic reflection 
can explain the broader Fe\,K profile in 2014, while in 2013 the more extended
 warm optically-thick corona hid most of the relativistic
reflection from the accretion disc, implying a smaller and narrower 
Fe\,K line. However, in contradiction with the expected viscous disc time-scale, 
we infer from the fit a significant increase in mass accretion rate through the disc
from $L_{\rm bol}/L_{\rm Edd}\sim 0.03$ (2013) to $0.07$ (2014) in only one year.  
This issue is discussed in section~\ref{sec:NT}.

Figure~\ref{fig:2013-2014-modelB} (bottom panel) displays the different intrinsic
(corrected for both Galactic reddenning and absorption) SED for both the 2013 (in red) and 2014 (in black) observations. The intrinsic luminosities from optical to hard X-rays are 1.07\,$\times$\,10$^{45}$\,erg\,s$^{-1}$ and 2.42\,$\times$\,10$^{45}$\,erg\,s$^{-1}$, respectively.
The outer disc emission (dotted curves) in 2014 (in black) is much stronger and peaks at a higher energy than in 2013 (in red) due to a  much smaller inner radius (that is, 14\,$R_{\rm g}$ and 85\,$R_{\rm g}$ in 2014 and 2013, respectively). 
 The UV band flux is significantly higher in 2014 than in 2013, as for example by a factor of about 50$\%$ for the UVW2 filter (\citealt{Lobban18}; see also Table~\ref{tab:OM}). This higher UV flux likely drives the requirement in the SED model for a lower inner disc radius ($R_{\rm corona}$) in 2014. 
The warm optically-thick Comptonisation components (dashed curves) also differ   
significantly between the two observations with different peak energies,
but nonetheless are the dominant process in both observations in the soft X-ray band below 1\,keV.   
The  hot Comptonisation component is much steeper in 2014
when the source is brighter as mentioned above, but becomes similar to the 2013 one above about 100\,keV.\\

In order to assess the impact on the spin value 
of the hypothesis of a full covering optically-thick corona, 
we relaxed the assumption where 
 $R_{\rm in}$ is tied to $R_{\rm corona}$. 
The inferred spin value of 0.86$^{+0.02}_{-0.01}$ is slightly higher than the spin value found assuming a full
covering warm optically-thick corona (that is, $R_{\rm in}$ tied to $R_{\rm corona}$), 
but still compatible within the error bars. 
For the 2014 observation, $R_{\rm corona}$ and $R_{\rm in }$ are found to be similar with 
$R_{\rm corona}$=12.0$^{+2.7}_{-3.1}$\,$R_{\rm g}$ and  
$R_{\rm in}$=13.9$^{+4.5}_{-3.8}$\,$R_{\rm g}$, while for the 2013
observation the two values differ significantly with 
$R_{\rm corona}$=77.3$^{+11.6}_{-8.5}$\,$R_{\rm g}$ and 
$R_{\rm in}$=14.2$^{+10.0}_{-3.7}$\,$R_{\rm g}$ ($\chi^{2}$/d.o.f.=3809.6/3515). 
For the 2013 observation, this suggests that beyond 14\,$R_{\rm g}$
some contribution of the reflection off the disc is observed and that the
optically-thick corona may be patchy above this radius.
This would be consistent with the rapid variability of the FeK emission complex, as discussed in \cite{Nardini16}. 
Alternatively, the reflection continuum from more distant material 
 (e.g. the outer disc, dense BLR clouds or the inner torus)    
may also become more important in the lower flux 2013 observation.

\section{Discussion}\label{sec:discussion}

\subsection{Impacts of the assumed fixed parameter values}

The disc-Comptonisation efficiency method using {\sc optxagnf/optxconv} 
has been applied to some other AGN 
\citep{Done12,Done13,Done16} but the spin value was either
fixed or almost unconstrained. 
 Indeed, this method requires a rather precise knowledge  
 of: the BH mass, the AGN distance and the accretion disc inclination angle
along the line of sight. Therefore, in practice, the uncertainties in the spin estimates are dominated
by their systematic uncertainties. 
As previously mentioned in section~\ref{sec:analysis}, the disc inclination angle 
($\theta$\,$\sim$30\,$^\circ$) is well determined here  
thanks to the mildly relativistic reflection component observed in the 2014 observation, 
and is found to be similar to that of the host galaxy \citep[that is, 26$^{\circ}$;][]{Nordgren95}. 
Therefore, we now investigate in this section
the impact of the assumed values of the fixed parameters: $E$(B$-$V),
the AGN distance, and the BH mass. 

\subsubsection{Impact of the assumed $E$(B$-$V) and of the AGN distance values}

We first investigated how the fit measurement depends on the assumption on the
values of the Galactic extinction ($E$(B$-$V)), and of the AGN distance. 
We allowed each of these to be free sequentially, and compared the inferred spin value with 
that found in the previous section. 

We assumed $E$(B$-$V)=0.128 (instead of 0.113) which   
corresponds to the original value from \cite{Schlegel98} without the
recalibration performed by \cite{Schlafly11}. 
We find a slightly lower value of the spin 
(0.79$\pm$0.05, Table~\ref{tab:optxconv}, column 2) compared to
that found previously with $E$(B$-$V)=0.113 (0.83$^{+0.05}_{-0.03}$, Table~\ref{tab:optxconv}, column 1), but still compatible within the error bars.
Indeed, for a given observed optical-UV luminosity a higher reddening along the line of sight implies a higher intrinsic luminosity, 
then a larger accretion rate (Eq.~\ref{eq:Lband}) 
and at last a smaller accretion efficiency (Eq.~\ref{eq:Lbol}). 

We then evaluated the impact on the fit results with a different AGN
distance of $D$=137.2\,Mpc (instead of 143.5\,Mpc) corresponding to the distance 
assuming the older cosmological constants from the five-year {\sl WMAP} 
\citep{Spergel03} where the corresponding cosmological constants 
  are $H_{\rm 0}$=71\,km\,s$^{-1}$\,Mpc$^{-1}$,
and $\Omega_{\Lambda_{\rm 0}}=0.73$. 
We find a negligible impact on the parameters fit compared to the
reference model (Table~\ref{tab:optxconv}, column 3).  
 
\subsubsection{Black hole mass value}\label{sec:mass}
Ark\,120 is one of the about 60 AGN \citep{Bentz15} for which the BH mass
has been determined via reverberation mapping \citep{Blandfor82,Peterson93}, 
which is one of the most reliable 
and direct methods to measure it in AGN
\cite[e.g.][and references therein]{Peterson04,Peterson14}. 
 
This method is based on the response of the broad emission lines of
the BLR to changes in the continuum.  
The virial BH mass is then estimated as $f$ ($\Delta$V$^{2}$\,$R_{\rm BLR}$/G), 
where $\Delta$V is the line width, $R_{\rm BLR}$ is the reverberation
radius, and $G$ is the Gravitational constant. 
 The quantity in brackets is called the `virial product' and
is determined by two directly observable parameters ($\Delta$V and $R_{\rm BLR}$). 
$f$ is a dimensionless factor -- often called the `virial factor' -- to take into
account the unknown BLR properties (structure, geometry, kinematics and
inclination with respect to the observer) and can be different from object to object.
A mean $f$, $\langle f \rangle$, value is currently determined via the $M_{\rm BH}$--$\sigma_{*}$
relationship assuming that it is the same 
for quiescent and active galaxies \citep{Ferrarese00,Gebhardt00,Woo13}, and by normalising the reverberation mapped AGN to
this relation (see \citealt{Peterson04} for a detailed explanation).     
However, this $\langle f \rangle$ quantity is not straightforward to infer since it
may, for example, depend on the bulge classification and/or on the presence of
 bars \citep[e.g.][]{Ho14,Graham11}, but see \cite{Graham14} on caveats
about bulge classification. 
The estimated values of $\langle f \rangle$ broadly range from about 4 and 6 
\citep[e.g.][]{Onken04,Shen08,Woo10,Park12,Grier13,Woo13,McConnell13,Ho14,Batiste17}.
So we now consider the minimum and maximum values of $\langle f \rangle$ reported
in the literature since the last few years.  
\cite{Ho14} find $\langle f \rangle$=6.3$\pm$1.5 for classical bulge galaxies such as
Ark\,120, 
 though \cite{Batiste17} find that their best-fit
relationship is insensitive to galaxy morphology. 
Alternatively, \cite{Grier13} find a smaller mean virial
factor with $\langle f \rangle$=4.31$\pm$1.05. This is the default value taken for the
calculation of the BH mass in `The AGN Black
  Hole Mass Database' \citep{Bentz15},\footnote{http://www.astro.gsu.edu/AGNmass/}  
using the calibration of the $M_{\rm BH}$--$\sigma_{\star}$ relation for AGNs based on
high-luminosity quasars hosts. 

The commonly used BH mass in the literature for Ark\,120 is 1.50$\times$10$^{8}$\,M$_{\odot}$. 
This is calculated from the virial product of 2.72$\times$10$^{7}$\,M$_{\odot}$ determined by \cite{Peterson04}, 
and assuming the viral factor of 5.5$\pm$1.8 from \cite{Onken04} (see Table~6 in \citealt{Peterson04}). 
So we now appraise the impact on the inferred
spin value still based on the virial product value from \cite{Peterson04} but using the 
lowest and highest virial product values as discussed above, 
that is, $\langle f \rangle$=4.31$\pm$1.05 \citep{Grier13} and  
$\langle f \rangle$=6.3$\pm$1.5 \citep{Ho14}. 
These values would correspond to BH masses of 1.17$\times$10$^{8}$\,M$_{\odot}$ and 1.71$\times$10$^{8}$\,M$_{\odot}$, respectively. 

As reported in Table~\ref{tab:optxconv}, the value of the
BH mass has the most important impact on the inferred spin value, 
compared to the $E$(B$-$V) and AGN distance values for which the impact is much less or even marginal. 
We find a spin value of 0.68$^{+0.05}_{-0.06}$ and 0.89$^{+0.04}_{-0.03}$ 
for $\langle f \rangle$=4.31$\pm$1.05 and  $\langle f \rangle$=6.3$\pm$1.5, respectively.
The higher the BH mass, the higher the BH spin value. 
Indeed, for a given observed $L_{\rm opt-UV}$, increasing/decreasing $M_{\rm BH}$ leads to a lower/higher $\dot{M}$ value (Eq.~\ref{eq:Lband}). Therefore, to reproduce the overall $L_{\rm bol}$ a higher/lower efficiency ($\eta$) is required (Eq.~\ref{eq:Lbol}) which corresponds to a higher/smaller BH spin value.
 
It is noteworthy that applying a general relativistic accretion disc corona model
-- but excluding the soft X-ray data (soft X-ray excess) --
to five higher accretion rate AGN (0.3\,$\lesssim$\,$\dot{M}$\,$\lesssim$\,0.5),
\cite{YouB16} show that the spin can be well constrained
 if the mass measurement is known to within 50\% accuracy (see also \citealt{Czerny11b,Done13}).
 Here, considering the lowest and highest BH masses for Ark\,120 as determined above, 
 this correspond to an accuracy of the BH value compared to the reference one of 22\% and 14\%, respectively. 
 
Other direct methods to determine the mass exist and have been applied to 
Ark\,120. Recently, \cite{Denissyuk15} used the radial velocities of 
emission features to infer a BH mass for Ark\,120 of
1.675$\pm$0.028$\times$10$^{8}$\,M$_{\odot}$, similar to that
found by \cite{Peterson04} and \cite{Ho15}. Another method is based on
polarisation of the broad emission lines \citep[e.g.][]{Afanasiev15,Songsheng18,Savic18},  
and has been applied to Ark\,120 by \cite{Afanasiev15}. The inferred a BH mass of
1.04$^{+1.38}_{-0.58}\times$10$^{8}$\,M$_{\odot}$ that is compatible
within its error bars with the values used in this work. \\

Finally, we estimate from the SED shape 
the required BH mass that would correspond to a non-rotating BH ($a$=0) 
or a maximally rotating BH ($a$=0.998). 
We find $M_{\rm BH}$=5.82$^{+1.33}_{-0.96}$~$\times$~10$^7$\,M$_{\odot}$ ($\chi^{2}$/d.o.f.=3820.3/3517), 
and $M_{\rm BH}$=3.34$^{+0.07}_{-0.26}$~$\times$~10$^8$\,M$_{\odot}$
($\chi^{2}$/d.o.f.=3850.7/3517), respectively.  
These would correspond, respectively, to mean virial factors of 
about 2.1 and 12.3 which are very discrepant from the mean values
published during the last few years. 
So, unless the individual virial factor for Ark\,120 strongly differs 
from the mean value reported for AGN,
these two `extreme' solutions appear unlikely. Therefore, 
an intermediate spin value of about 0.7--0.9 is strongly favoured for Ark\,120.

\subsection{Deviations from Novikov-Thorne thin disc ?}\label{sec:NT}

As shown by general relativistic magnetohydrodynamic (GRMHD)
simulations of accretion onto stellar BHs, contrary to the basic
assumptions of the thin disc model of \cite{Novikov73} (NT), there can be significant
magnetic stress throughout the plunging region. This means that
additional dissipation and radiation can be expected.  
However, as shown by \cite{Kulkarni11} for an accretion rate of 0.1 times Eddington 
and an accretion disc inclination angle of 30$^{\circ}$, 
that is,\ similar to the values found for Ark\,120,
the discrepancies on the inferred spin values 
between calculations from GRMHD simulations and a NT disc assumption are only 
about 0.007 and 0.02 for spin values of 0.9 and 0.7, respectively \citep[see also,][]{Penna10,Noble11,Penna12,Zhu12,Sadowski16}. 
This would mean that for objects with both low-to-moderate
accretion rates and inclination angles, as Ark\,120, 
 such systematic error is marginal (and much smaller than the error bars from spectral measurements), and that the NT disc theory is adequate and can be safely applied.  

However, for the Ark\,120 the warm and hot Comptonisation components are dominant in the SED. Therefore, the hot corona in particular may well have 
significant scale height, especially during the 2013 low-flux state observation that would correspond 
to a low accretion rate ($L/L_{\rm Edd}$) of about 0.03 and may be in form of a hot inner flow with large scale height  
\citep[][see their Fig.\ 2 for the disc-corona scheme]{Kubota18}. 
This could mean that the flow has some pressure support so is sub-Keplerian, and then that advection is important
as a cooling mechanism, which acts to suppress $\eta$ below that expected from a thin disc \citep{Narayan95}.
Therefore, dedicated calculations  are required to determine possible deviations from the NT model
(and then ultimately on spin measurements) for moderate accretion rate AGN for which warm and hot Comptonisation are the dominant processes. 
However, it is noteworthy that \cite{YouB16} from their general relativistic accretion disc-corona  model
find that for a BH mass of 10$^{8}$\,M$_{\odot}$ and spin of 0.9 the disc thickness (H/R) is much less
than 0.1 (see their Fig.\ 13) for an accretion rate value of about 0.08 of Eddington, that is, similar to that of Ark\,120 in 2014.\\

Another clear difference for Ark\,120 with the NT thin disc predictions is the variability. 
From the best simultaneous fit of the 2013 and 2014 observations, we infer a significant
increase in mass accretion rate through the disc from 0.03 to 0.07\,$L_{\rm bol}$/L$_{\rm Edd}$ in only one year, but a standard thin disc around a SMBH cannot vary on such timescale. Indeed, the radial mass accretion rate change via viscous processes has a time-scale of $t_{\rm orb}(R)/[\alpha (H/R)^2]$. For Ark,120, assuming an orbital time-scale at $R$=100\,$R_{g}$, a viscosity parameter of 0.1 and a H/R (H is the height scale of the disc) value of 0.1, 
this corresponds to $\sim$\,150 years.  
Moreover,  the optical-UV flux significantly changes in less than a year, varying for example by a factor of 50$\%$ in the UVW2 band between 2013 and 2014 (e.g. see Fig.~\ref{fig:swift}, \citealt{Lobban18}).
Similar rapid changes in the optical-UV flux are typically seen in other BLS1s,
especially those at low $L_{\rm bol}/L_{\rm Edd}$ \cite[e.g.][]{Macleod10,Kozlowski16,Simm16,Rakshit17}.
These are generally assumed to be from reprocessing, where the X-ray flux illuminates the outer disc (e.g. \citealt{Buisson17}),
and adds to the intrinsic emission. An additional reprocessed component in the optical-UV would lead us to 
overestimate the value of the intrinsic $\dot{M}$, so to an underestimate of BH spin via the efficiency argument
\citep{Kubota18}. 
Nonetheless, changes as large as about 50$\%$ in the UV flux are unlikely to be driven by X-ray reprocessing, as the UV flux in Ark\,120 is much higher than in the X-ray band.  
Besides, detailed models of the expected optical-UV variability from X-ray reprocessing  fail 
to fit the excellent long term simultaneous optical-UV-X-ray datasets, 
and would imply a larger disc size than expected by standard thin disc 
(e.g. \object{NGC\,5548}: \citealt{McHardy14,Edelson15,Gardner17}; 
\object{NGC\,4151}: \citealt{Edelson17}; Ark\,120: \citealt{Gliozzi17}; \object{Fairall\,9}: \citealt{Pal17}; \object{NGC4593}: \citealt{Cackett18,Pal18}; Microlensing studies: \citealt{Morgan10,Dai10}).  
Such large discs should be significantly brighter than observed and this discrepancy may be explained for example by a flatter temperature profile than in NT discs, from scattering of a significant part of the optical flux on larger scales, by electron scattering in the disc atmosphere \citep{Dai10,Morgan10,Hall18}. 

To resolve this issue, \cite{Gardner17} propose an alternative scenario where  
the observed optical-UV lags do not arise from hard X-ray reprocessing 
of the accretion disc emission from the hot corona, because it is shielded 
by the soft X-ray excess region, but instead arise from
reprocessing of the far-UV emission by optically thick clouds in the
inner regions of the BLR (see their Fig.\ 14). 
In addition, \cite{Lawrence18} argue that reprocessing by "clouds lifted out of the disc" might solve this "viscosity crisis".  
Moreover, as suggested by \cite{Cackett18} studying NGC\,4593, diffuse emission 
from the BLR must also contribute significantly to the interband lags (see also, \citealt{McHardy18,Lawther18}).

It has been also proposed by \cite{Noda18} that this faster than expected disc emission variability may be connected to the propation of heating-cooling waves across the disc which can change the accretion rate locally on faster timescale than the viscous one. Such changes mean that the flow is non-stationary with the accretion rate, not always constant with radius as assumed in the NT model. In addition, magnetically elevated discs, which are thicker than NT disc, would lead to much shorter inflow times \citep{Dexter19}. 

All these proposed scenarios show the importance for intensive multi-wavelength monitorings (from short to long timescales) of accretion disc emissions to better understand their origin(s) and check for genuine possible departure(s) from the NT model in AGN. However, the optical-UV to X-ray timing analysis of Ark\,120 showed that the time lag between X-rays and the U band is only about two days and the time lag between UVW2 is compatible with zero within the uncertainties \citep{Lobban18}. Therefore, these timescales are similar to duration of the multi-wavelength simultaneous observations used in this work. Moreover on such short timescales, the variability of the optical and UV bands is within only a few percent, with a maximum of six percent for the UVW2 band. Therefore, this restrained optical-UV variability should prevent from too large a departure from energy conservation in a steady state accretion disc as used in the optxagnf model. \\

In conclusion, we caution that our results
on BH spin depend on the detailed disc-corona structure, which is not yet fully constrained.   
However, the realistic parameter values (e.g. log($L_{\rm bol}$/L$_{\rm Edd}$) range, disc inclination angle)
found from this analysis for Ark\,120 seem to show that this is a promising method to determine spin in BLS1s.

\subsection{Other spin measurement methods from X-rays} 

One way to check the disc-Comptonisation efficiency method used in this work and then the disc-corona structure assumed, is to compare the inferred spin value with those derived from other methods, when applicable.

\subsubsection{X-ray relativistic reflection fitting}

A promising way to strengthen the spin determination would be to compare the spin determination 
for the same AGN from the disc-Comptonisation efficiency method
 and from relativistic reflection modelling. 
For this, finding AGN -- with well constrained BH mass -- 
displaying at different periods spectra dominated
by either Comptonisation or relativistic reflection would be of great interest. 
 This could be the case for Ark\,120. Indeed, the 2007 {\sl Suzaku} spectrum -- with an X-ray flux between the ones observed in 2013 and 2014 -- displayed an apparently  broader FeK$\alpha$ line compared to that observed in 2014 with {\sl XMM-Newton} \citep{Nardini16}, and the X-ray {\sl Suzaku} spectrum has been interpreted as due to relativistic reflection emission \citep{Nardini11}. 
However, the lack of good S/N data above 30\,keV from this 2007 {\sl Suzaku} observation precludes for a  discrimination between Comptonisation versus relativistic reflection as the dominant process (as shown in \citealt{Porquet18}).  

In the case where the spin values inferred from the two methods do not agree,
this will give us an indication that one or both models 
have to be improved until both inferred spin values match each other. 
Indeed, there are a lot of caveats in the determination of AGN spin, such as the presence of a strong warm absorber and/or wind component(s). 
For example, such a comparison between X-ray relativistic reflection fitting and the disc-Comptonisation efficiency
method has been applied to the `complex' NLS1 \object{1H 0707--495} by \cite{Done16}. Fixing the parameters inferred
from relativistic reflection models (that is high spin, moderate inclination and low-mass BH) to the efficiency method,
\cite{Done16} infer a non physical extreme accretion rate value of 140--260 times the Eddington limit for this object.
Therefore, they argue that strong winds expected for such type of objects could bias the spin measurement in relativistic reflections models. 
 Indeed, a strong wind could alter the determination of the spin if the sharp drop usually interpreted as the blue wing of a relativistic FeK line is actually due (at least in part) to a blueshifted absorption feature(s) \citep{Kosec18,Parker18b,Jiang18}. 
This could also suggest that 
for high-accretion rate AGN like 1H 0707--495 pure relativistic reflection modelling is not adequate. 
Indeed, as shown in  \cite{Kammoun18} the spin can be recovered even if complex warm and cold absorptions are present (provided that both reflection is strong and the spin is high, assuming a lamp-post geometry), but leaving such effects unmodelled can introduce significant and poorly controlled systematics. 

Moreover, before applying one of these fitting methods the origin of the soft excess must be robustly determined
  to avoid biased determination of the spin value \citep{Patrick11,Boissay16,Porquet18}.
  For example, \cite{Walton13} applying a pure relativistic reflection model to the NLS1 \object{Ton\,S180} spectrum infer a high spin value of about 0.9 and a very high emissivity index ($q>$8) to be able to reproduce the extremely smooth soft X-ray excess. 
  This was contradicted recently by \cite{Parker18a} (see also discussion in \citealt{Nardini12}) who -- applying a state-of-the-art relativistic reflection model ({\sc relxill$\_$lp}; \citealt{Dauser10}) to a higher S/N X-ray spectrum --
  ascertain that the soft X-ray excess in Ton\,S180 cannot be accounted for by reflection and inferred from the 3--10\,keV energy band a low spin value (a$<$0.4).
  As for Ark\,120 \citep{Porquet18}, they find for the broad-band spectrum of Ton\,S180 can be modelled by a two-component Comptonisation continuum plus mildly relativistic disc reflection component. 
This emphasises the need to describe the broad band continuum correctly in order to reliably estimate the black hole spin.

\subsubsection{X-ray QPOs and X-ray polarisation} 

As discussed in the introduction section there are two other methods to determine
spin from X-ray data: high-frequency QPOs and polarisation. 

High-frequency QPO are primarily dependent on the BH 
mass and spin  \citep{Abramowicz01,Remillard06}.  
However, these 
have only recently been detected in AGN as for example in \object{RE\,J1034+396} \citep{Gierlinski08a,Alston14}, in \object{MS\,2254.9-3712}
\citep{Alston15}, in \object{2XMM\,J123103.2+110648} \citep{Lin13}, and
possibly in 1H 0707-495 \citep{Pan16}. These are all high accretion rate
objects, with $L_{\rm bol}/L_{Edd}$ $\gtrsim$ 1, which are in a very different accretion regime compared to Ark 120. 
Additionally, transient QPOs have been detected in tidal disruption events
 as in \object{Swift\,J164449.3+573451} \citep{Reis12,Wang14}.
So far, the only results from all these detections are a single upper limit of about 0.08 for the BH spin 
for RE\,J1034+396 \citep{Mohan14},
though again it depends on a good BH mass estimate which is lacking in this AGN \citep{Czerny16}.

Special and general relativistic effects strongly modify the
polarisation properties of the radiation observed at infinity.
For instance, a spin-dependent rotation of the polarisation angle
with energy is expected for the disc thermal emission
\citep[e.g.][]{Stark77,Dovciak08,Li09,Schnittman09}. 
For a stellar-mass accreting black hole in the soft state, this rotation
happens mostly above 1 keV, and can therefore be searched for
by the X-ray polarimetry
missions already approved (IXPE, \citealt{Weisskopf16}) or under
study (eXTP, \citealt{Zhang16}), which work in the 2$-$8 keV
band. In AGN, this energy band is dominated by coronal emission
and reflection, and the use of polarimetry to derive the black hole
spin is still possible but less straightforward 
\citep{Schnittman10,Dovciak11,Beheshtipour17,Marin18,Marin18b}, 
not mentioning further complications arising
from the contribution of reflection from 
parsec-scale AGN components such as the molecular
torus. Effects in different coronal geometries, like the
one adopted in this paper, are yet to be investigated to predict
whether the spin and mass of the central SMBH can be
robustly extracted
from the polarimetric signal of AGN in the IXPE band.

\section{Conclusion}\label{sec:conclusion}

We performed optical to hard X-ray SED fitting of Ark\,120, using {\sl XMM-Newton} and {\sl NuSTAR} data obtained in March 2014 and
February 2013 applying the disc-Comptonisation efficiency method. 
We used the {\sc optxconv} model (based on the {\sc optxagnf} model, \citealt{Done13})
 to self consistently model the outer disc emission and the inner (warm and hot) Comptonisation components, 
     taking into account both inclination and relativistic effects. For
     the relativistic reflection we used as incident spectrum the
     Comptonisation spectral shape for self-consistency.  
Assuming a full covering optically-thick corona for the warm 
     Comptonisation component, meaning that any relativistic
     reflection below the optically-thick corona radius $R_{\rm corona}$ is hidden to
     the observer, we find a good fit for both datasets, though some excess at high energy indicates that the hot Comptonisation component has a temperature larger than 100\,keV. 

We find that the warm optically-thick corona ($\tau$\,$\sim$\,9--12) has a  
  temperature ($kT_{\rm e}$\,$\sim$0.4--0.5\,keV) that slightly increases when
  the source is brighter (2014 March 22). We also confirm the softer when
  brighter behaviour for Ark\,120 reported previously
\citep{Matt14,Gliozzi17,Lobban18}.

  We infer a receding coronal radius (at a statistical significance of about 5.5$\sigma$) with increasing flux (by about a factor of two) 
  from about 85$^{+13}_{-10}$\,$R_{\rm g}$ 
   to about 14$\pm$3\,$R_{\rm g}$ from February 2013 to March 2014. 
   However, there is some 
   indication that for the 2013 observation the optically-thick corona
 may be patchy above about 14\,$R_{\rm g}$. 
We find a well constrained spin value of 0.83$^{+0.05}_{-0.03}$ assuming a 
   reverberation BH mass of 1.50$\times$10$^{8}$\,M$_{\odot}$. 
We investigated the impact of assumption of the $E$(B$-$V) values as well
as the SMBH properties (distance, mass). We find that the most
important impact on the spin value is due to the BH mass. 
However, we are able to infer that the likely SMBH spin is located
between 0.68$^{+0.05}_{-0.06}$ and 0.89$^{+0.04}_{-0.02}$ even if we
assumed the minimum and maximum mean virial factor ($\langle f \rangle$) values reported
in the literature.

In conclusion, for the first time we are able to infer tight constraints on the spin of a SMBH 
using the disc efficiency method, via modelling the optical-UV to X-ray emission from the Comptonised disc spectrum. 
This was possible thanks to the properties of Ark\,120, namely that it is a bare AGN devoid of intrinsic absorption, its reliable black hole mass determined via reverberation mapping, as well as the presence of a mild relativistic reflection in the 2014 observation.
 The latter allows us to obtain a good constraint on the inclination angle, another important parameter to constrain the spin.

 However, this method crucially depends upon the assumption that 
the emissivity is given by the expected thin disc Novikov-Thorne relation.
Indeed, it is possible that the accretion flow
in Ark 120 does not comply with the predictions of a thin disc, firstly as its X-ray spectrum is dominated by
warm and hot Comptonisation components rather than by the disc emission, and secondly as
the optical-UV varies on much faster time-scales than expected, as well as the accretion rate. 
 Therefore, a much better understanding of the accretion flow
is required before this technique to measure the spin can be deemed as robust. 
In this framework, a comparison of BH
spin measurements using different methods in X-rays (e.g. relativistic reflection modelling, high-frequency QPO,
and polarisation properties that will be accessible in the near future) will be of great interest. 
However, the realistic parameter values, such as the Eddington ratio and the accretion disc inclination angle, 
found from this method for Ark\,120 suggest that this is a promising technique to determine spin in BLS1s.\\



\begin{acknowledgements}
We  thank  the  anonymous  referee  for  her/his  comments. 
Based on observations obtained with the {\sl XMM-Newton}, and ESA science
mission with instruments and contributions directly funded by ESA
member states and the USA (NASA). 
 This work made use of data from the
{\sl NuSTAR} mission, a project led by the California Institute of
Technology, managed by the Jet Propulsion Laboratory, and
funded by NASA. 
This research has made use of the {\sl NuSTAR} 
Data Analysis Software (NuSTARDAS) jointly developed by
the ASI Science Data Center and the California Institute of Technology.
J.\ N.\ Reeves acknowledges support through NASA grant NNX15AF12G.  
A.\ L.\ acknowledges support from the UK STFC under grant ST/M001040/1. 
E.\ N.\ acknowledges funding from the European Union’s Horizon 2020 research and innovation programme under the Marie
Sklodowska-Curie grant agreement No.\ 664931. 
F.\ M.\ is grateful to the Centre national d’études spatiales (CNES) and its post-doctoral 
grant “Probing the geometry and physics of active galactic nuclei with ultraviolet and 
X-ray polarized radiative transfer". C.\ R. acknowledges support from the CONICYT+PAI 
Convocatoria Nacional subvencion a instalacion en la academia convocatoria a\~{n}o 2017 PAI77170080 (C.R.).
\end{acknowledgements}

%
%


\end{document}